\documentclass[12pt,a4paper]{article}
\usepackage{amsmath}
\usepackage{amsthm}
\usepackage{amsfonts}
\usepackage{amssymb}
\usepackage{booktabs}
\usepackage{graphicx}
\usepackage{color}
\definecolor{darkgreen}{rgb}{0,0.5,0}
\definecolor{purple}{rgb}{1,0,1}
\newcount\Comments  %
\newcommand{\kibitz}[2]{\ifnum\Comments=1\textcolor{#1}{#2}\fi}

\newcommand{\ra}[1]{\renewcommand{\arraystretch}{#1}}
\numberwithin{equation}{section}
\usepackage{srcltx}
\usepackage{authblk}
\usepackage[left=2.00cm, right=2.00cm, top=2.00cm, bottom=2.00cm]{geometry}
\theoremstyle{definition}
\newtheorem{theor}{Theorem}[section]
\newtheorem{exmp}{Example}[section]
\begin{document}
\title{Why is my  rational Painlev\'e V solution not unique?}

\author[1]{ H. Aratyn}
\author[2]{J.F. Gomes} 
\author[2]{G.V. Lobo} 
\author[2]{A.H. Zimerman}
\affil[1]{
Department of Physics, 
University of Illinois at Chicago, 
845 W. Taylor St.
Chicago, Illinois 60607-7059, USA}
\affil[2]{
Instituto de F\'{\i}sica Te\'{o}rica-UNESP,
Rua Dr Bento Teobaldo Ferraz 271, Bloco II,
01140-070 S\~{a}o Paulo, Brazil}

\maketitle

\abstract{
Under special conditions the Painlev\'e V equation has more than one 
rational solution solving it with the same parameters.

In the setting of formalism that identifies 
points on orbits of the fundamental shift operators of $A^{(1)}_{3}$ affine 
Weyl group with rational solutions we derive conditions for 
such non-uniqueness to occur.

We identify the seed solutions from which the non-unique solutions are
generated and put forward a method to systematically obtain their closed
expressions from the underlying seed solutions.

}

\section{Background and Introduction}
We formulate our study of rational solutions of Painlev\'e V (PV) equation in  the setting 
of Hamiltonian formalism with the Hamiltonian
\begin{equation}
H= -q\,(q-z)\,p\, (p-z)+(1-\alpha_1-\alpha_3)\, p q +\alpha_1 z
p-\alpha_2 z q \, ,
\label{pqHam}
\end{equation}
where $\alpha_i, i=1,2,3$ are three constant parameters and
$q, p$ are two canonical variables that satisfy Hamilton equations 
given below in \eqref{hameqs}. These equations are invariant
under B\"acklund transformations, $\pi, s_i,i=1,{\ldots}, 4$, that
satisfy the $A^{(1)}_{3}$
extended affine Weyl group relations \cite{noumibk,Noumiwkb} 
shown in equation \eqref{fun.rel}.

Within the $A^{(1)}_{3}$
extended affine Weyl group one defines an abelian 
subgroup of translation operators also referred to
as  fundamental shift operators. The four translation operators, $T_i,
i=1,2,3,4$, are given by expression $T_i=r_{i+3} r_{i+2} r_{i+1} r_i$, 
where $r_i=r_{4+i}=s_i$ for $i=1,2,3$ and $r_4 = \pi$.

We will be working within the formalism 
presented in \cite{AGLZ}, which described construction of  rational solutions
of PV equation  out of actions of translation  operators
on seed solutions. 
Crucial for this construction is that rational solutions fall into two classes 
depending on which of the two types
of seed solutions,  $\left(q=z/2,p=z/2\right)_{\mathsf{a}}$
or $\left(q=z,p=0\right)_{\mathsf{a}}$, they have been derived
from by actions of the fundamental shift operators.
It was shown in  \cite{AGLZ} 
that this procedure %
reproduces all conditions \eqref{Ia}, \eqref{Ib}, \eqref{II},
\eqref{IIIa} and \eqref{IIIb} that are
satisfied by parameters of rational solutions of PV as first 
obtained and classified in \cite{kitaev}.

The discussion of seed solutions becomes more transparent
in a framework of  $N=4$ periodic dressing chain formulated in
\cite{AGLZ}, see also \cite{adler,veselov} for relation between
periodic dressing chains and Painlev\'e equations.   For one class of solutions the shift operators
are all allowed to act without any restriction. The underlying 
seed solution is characterized by having all components $j_i$ of an
underlying dressing chain being equal to $z/4$: $j_i=(z/4)(1,1,1,1)$.
In the framework of canonical variables of the Hamilton formalism such structure 
corresponds to  $\left(q=z/2,p=z/2\right)_{\mathsf{a}}$
, with the variable $\mathsf{a}$
determining the $\alpha_i$ parameter of PV equations as
$\alpha_i=(\mathsf{a}, 1-\mathsf{a}, \mathsf{a}, 1-\mathsf{a})$
\cite{AGLZ}. The rational solutions  resulting from action of
translation operators on this seed solution are Umemura polynomials. There is no
degeneracy in this sector (to each of the parameters
$(\alpha_1,\alpha_2,\alpha_3, \alpha_4)$ 
corresponds a unique solution) due to the fact that all the operators
that generate solutions are fully invertible. 

A remaining class of rational solutions is derived out of the seed solution 
having one negative $j_i$ component among the four components with
$\pi$ automorphism shifting the position  of the negative component
among the other components as shown below equation  \eqref{wata6}.
It is therefore sufficient to carry out our analysis for one such
configuration. In the setting of PV Hamiltonian formalism 
the corresponding seed solution is labeled by a
parameter $\mathsf{a}$ or $\mathsf{b}$ such that
$\left(q=z,p=0\right)_{\mathsf{a}}$ denotes a solution of PV equations
\eqref{hameqs} with the parameters $\alpha_i=(\mathsf{a},0,0,2-\mathsf{a})$.

In the latter case there are exclusion rules 
that dictate which of the shift operators or their inverse 
are allowed to act  on the seed solution.
These rules simply  exclude those operators that will produce divergencies. 
The corresponding rational solutions are special functions
and have parameters that satisfy general rules 
\eqref{IIIa} and \eqref{IIIb},
derived on basis of action by allowed translations on the
seed solutions $\left(q=z,p=0\right)_{\mathsf{a}}$.
The rules 
\eqref{IIIa} and \eqref{IIIb} are unchanged  by additional
transformations by the B\"acklund transformations $s_i, i=1,2,3,4$. In
that sense the main properties of the parameters of rational
solutions are derived solely from action of translation operators.

Degeneracy occurs for this latter type of rational solutions to PV
equation with  $\alpha_i$ parameters being shared by two different solutions.
The conditions for degeneracy are met when we can not equate 
expressions for solutions generated by action by translation
operators of the type:
\begin{equation}
Y_{m, \mathsf{b}} 
= T_2^{-m_2} T_4^{m_4} \left(q=z,p=0\right)_{\mathsf{b}},
\;\quad m_2, m_4 \in \mathbb{Z}_{+}\, ,
\label{Ymdef}
\end{equation}
with solutions constructed as 
\begin{equation}
X_{M, \mathsf{a}}  = M T_2^{-n_2} T_4^{n_4}\,
 (q=z,p=0)_{\mathsf{a}} , \;\quad \; n_2, n_4 \in \mathbb{Z}_{+}\, ,
\label{XMdef}
\end{equation}
with translation operators being augmented by additional 
B\"acklund transformations $s_i$ or their
products contained in expression $M$ to be determined from a number of
conditions including one that requires $X_{M, \mathsf{a}}$ to be
finite. $\mathbb{Z}_{+}$ is defined throughout the paper as
containing positive integers and zero while a symbol $\mathbb{Z}$
describes a set of all integers.

We will determine $M$ for which an attempt to equate the above two 
solutions : $Y_{m, \mathsf{b}}  =X_{M, \mathsf{a}}$, with a common 
$\alpha_i$, fails because it would lead to an infinity.
Such divergence is a precursor of degeneracy because
it prevents the two different expressions for rational solutions from being %
equal although they solve PV equation with identical
parameters. 

The (two-fold) degeneracy will occur for the $\alpha_i$ parameters given by
\begin{equation} \begin{split}
\alpha_i^{(1,2)} &= T_2^{-m_2} T_4^{n_2+n_4}
(\mathsf{b},0,0,2-\mathsf{b}),\quad
\mathsf{b}=2 n_2, \;\, n_2 \in
\mathbb{Z}\;  \& \; n_2 >0,
\\
&= 2(n_2+m_2,-m_2,-n_2-n_4,1+n_4) \;
\quad \; m_2, n_4 \in \mathbb{Z}_{+}, \; 
\label{alpha12}
\end{split}
\end{equation}
and corresponding  to solutions \eqref{Ymdef} and \eqref{XMdef}
with $M=s_1 s_2$  and  integer $m_4=n_2+n_4$.

Another independent class of parameters with a two-fold degeneracy corresponds to
$M=\pi s_1$ and is given by 
\begin{equation} \begin{split}
\alpha_i^{(1)} &= T_3^{m_2} T_1^{n_4} T_4^{n_2}
(\mathsf{a},0,0,2-\mathsf{a}),\quad
\mathsf{a}=2+2 n_2, \, n_2 \in
\mathbb{Z}\;  \& \; n_2 > m_2 \ge 0,
\\
&= 2(1+n_2+n_4,-m_2,m_2-n_2,-n_4) \;
\quad \; m_2, n_4 \in \mathbb{Z}_{+}\, .
\label{alphapi1}
\end{split}
\end{equation}
Here $T_3 $ acts formally on $\alpha_i$ while it is not allowed to
act on a seed solution $\left(q=z,p=0\right)_{\mathsf{a}}$.
Since $\pi s_1 = s_3 s_4 T_1^{-1}$, as it will be shown below,
we could have alternatively chosen $M= s_3s_4$
to generate this class of solutions. 

In section \ref{hamsym}, we will introduce the PV equation invariant
under $A^{(1)}_{3}$ extended affine Weyl group of its B\"acklund transformations.
The abelian subgroup of $A^{(1)}_{3}$ consisting of 
translation operators is constructed in subsection \ref{translations},
where  we  also give details on their actions on the
parameters of solutions as well as their commutation relations with
B\"acklund transformations. %
The seed solutions  of PV equation and actions
of translation operators on these solutions
are shown in subsection \ref{seeds}.
The parameters of rational solutions of PV aequation re described in subsection
\ref{classification} by constructing these rational solution out of
the seed solution through actions of translations.
The special case of solutions constructed from the seed solutions
\eqref{solution2} is free of degeneracy as all translations and
B\''acklund symmetries are invertible. This is discussed in subsection
\ref{subsection:item1&2} together with another related feature that 
action of $s_i$ B\"acklund symmetries can, for special value of the parameters,
be represented by action
of translation operators acting on this class of solutions.

Conditions for degeneracy are derived in section \ref{vinculos}
and the corresponding solutions are explicitly constructed in subsection
\ref{subsection:realization}
The two-fold character of degeneracy for parameters shown
in equation \eqref{alphapi1} is proven in subsection \ref{subsection:x1x4}
 with explicit examples of such solutions given in subsection
 \ref{subsection:s3s4T2T4}.
Examples of solutions and the degeneracy 
corresponding to parameters $\alpha_i^{(1,2)}$ defined in \eqref{alpha12}
is given in subsection \ref{example:s1s2}.

In section \ref{discussion} we offer few concluding remarks and
indicate directions for  future research.

\section{The Hamiltonian model,  its B\"acklund symmetries and PV
equation}
\label{hamsym}
We work with the Hamiltonian formalism with two canonical variables $q, p$ 
depending on coordinate $z$, that
satisfy the following Hamilton equations:
\begin{equation}\begin{split}
z q_z &= -q(q-z)(2p-z)+(1-\alpha_1-\alpha_3)  q +\alpha_1 z \, ,\\
z p_z &= p (p-z)(2q-z) - (1-\alpha_1-\alpha_3) p +\alpha_2 z \, ,
\label{hameqs}
\end{split}
\end{equation}
derived as Hamilton
equations, $z q_z = {d H}/{d p}$ , $z p_z =- {d H}/{d q}$
from the Hamilton function $H$ defined in \eqref{pqHam}.

Equations \eqref{hameqs} are invariant under  B\"acklund transformations shown in Table \ref{backsymham},
\begin{table}[h]
	\centering
	\ra{1.8}
	\begin{tabular}{ccccccc}
		\toprule
		& $q$ & $p$ & $\alpha_1$ & $\alpha_2$ & $\alpha_3$ & $\alpha_4$ \\
		\midrule
		$s_1$ & $q$ & $p + \frac{\alpha_1}{q}$ & $-\alpha_1$ & $\alpha_1 + \alpha_2$ & $\alpha_3$ & $\alpha_1 + \alpha_4$ \\
		$s_2$ & $q- \frac{\alpha_2}{p}$ & $p$ & $\alpha_1 + \alpha_2$ & $-\alpha_2$ & $\alpha_2 + \alpha_3$ & $\alpha_4$ \\
		$s_3$ & $q$ & $p -\frac{\alpha_3}{z-q}$ & $\alpha_1$ & $\alpha_2+\alpha_3$ & $- \alpha_3$ & $\alpha_3 + \alpha_4$ \\
		$s_4$ & $q + \frac{\alpha_4}{z-p}$ & $p$ & $\alpha_1 + \alpha_4$ & $\alpha_2$ & $\alpha_3 + \alpha_4$ & $-\alpha_4$ \\
		$\pi$ & $z-p$ & $q$ & $\alpha_4$ & $\alpha_1$ & $\alpha_2$ & $\alpha_3$ \\
		\bottomrule
	\end{tabular}
	\caption{$A^{(1)}_{3}$ B\"acklund transformations}
	\label{backsymham}
\end{table}
with a new parameter $\alpha_4$  defined as 
$\alpha_4=2- \alpha_1-\alpha_2 -\alpha_3$.

The symmetry generators $\pi, s_i,i=1,{\ldots}, 4$ satisfy the following 
fundamental relations 
\cite{adler,Noumiwkb,noumibk}:
\begin{xalignat}{2}
   s_i^2=1, &\qquad \quad s_i s_j =s_j s_i~(j \ne i,i \pm 1), &
   s_i s_j s_i =s_j s_i s_j~(j=i \pm 1),\nonumber \\
  \pi^4=1, & \qquad \quad \pi s_j =s_{j+1}\pi.&
\label{fun.rel}
\end{xalignat}
of the extended affine Weyl group $A^{(1)}_{3}$.

In terms of  $y={w}/(w-1)$, where $w = {q}/{z}$, we can cast 
the Hamilton equations \eqref{hameqs}
as a second order nonlinear differential
equation known as  Painlev\'e V equation :
\begin{equation} 
y_{t t}
= -\frac{y_t}{t}+\left( \frac{1}{2y}+ \frac{1}{y-1}\right)
y_{t}^2 + \frac{(y-1)^2}{t^2} \left({ \alpha} y + 
\frac{{ \beta}}{y} \right) + \frac{{ \gamma}}{ t} y + { \delta} 
\frac{y (y+1)}{y-1}\, ,
\label{finalPV}
\end{equation}
where we changed a variable $z \to t$ with 
$t = \epsilon z^2/2$ and where 
\begin{equation} 
{ \alpha}= \frac18 \alpha_3^2,\;\; { \beta}=- \frac18 \alpha_1^2,\;\;
{ \gamma}= - \frac{1}{2 \epsilon }
( 2-2 \alpha_2-\alpha_1-\alpha_3)= \frac{\alpha_2-\alpha_4}{2 \epsilon },\;\;
{ \delta}=   -\frac12 \frac{1}{\epsilon^2}\, .
\label{paramPVbar}
\end{equation}
For ${ \delta}$ to take a conventional value of $-\frac12$ we need 
$ \epsilon^2 = 1$.

\subsection{Translations within the extended affine Weyl group ${A^{(1)}_{3}}$}
\label{translations}
The extended affine Weyl group ${A^{(1)}_{3}}$ contains an abelian 
subgroup of fundamental shift operators $T_1, {\ldots} ,T_4$ :
\begin{equation}
T_1=\pi s_3s_2s_1, \quad T_2=s_1\pi s_3s_2, \quad T_3=s_2s_1\pi s_3, 
\quad T_4=s_3s_2s_1\pi, 
\label{trans}
\end{equation}
that commute among themselves 
\[ T_i T_j =T_j T_i\, .\]
The operators $T_i, {\ldots} ,T_4$  generate the following translations 
when acting on the parameters of PV equation:
\begin{equation}
\begin{split}
T_1  (\alpha_1, \alpha_2, \alpha_3, \alpha_4) &=
 (\alpha_1+2, \alpha_2, \alpha_3, \alpha_4-2) \, ,\\
T_2  (\alpha_1, \alpha_2, \alpha_3, \alpha_4) &=
 (\alpha_1-2, \alpha_2+2, \alpha_3, \alpha_4) \, ,\\
T_3  (\alpha_1, \alpha_2, \alpha_3, \alpha_4) &=
 (\alpha_1, \alpha_2-2, \alpha_3+2, \alpha_4) \, , \\
 T_4  (\alpha_1, \alpha_2, \alpha_3, \alpha_4) &=
 (\alpha_1, \alpha_2, \alpha_3-2, \alpha_4+2) \, ,
\label{Tialpha}
\end{split}
\end{equation}
The inverse $T_i^{-1}$ operators are
\begin{equation}
	T_1^{-1} =s_1s_2s_3 \pi^3, \quad T_2^{-1}=s_2 s_3 \pi^3 s_1, \quad T_3^{-1}=s_3\pi^3 s_1 s_2, 
	\quad T_4^{-1} =\pi^3 s_1s_2s_3\, .
\label{inverseT}
\end{equation}
As follows from their definitions \eqref{trans},  the translation
operators satisfy the following commutation relations 
\begin{equation} \begin{split}
s_i T_i&=T_{i+1}s_i,\,\; s_i T_{i+1}= T_i s_i, \quad s_i T_j=T_j s_i,
\quad  \pi\; T_i =T_{i+1}\, \pi\\
s_i T_i^{-1}&=T_{i+1}^{-1}s_i,\,\; s_i T_{i+1}^{-1}= T_i^{-1} s_i,
\quad s_i T_j^{-1}=T_j^{-1} s_i,\;j\ne i, i+1 \, ,
\end{split}
\label{lemmasi}
\end{equation}
with B\"acklund transformations $s_i$ for $i=1,2,3,4$ and automorphism
$\pi$.

In addition, we also find the following identities
for products of neighboring $s_i$ that reduce to one single
$s_i$ multiplied by the shift operator and an automorphism $\pi$:
\begin{equation}
\begin{split}
s_i s_{i+1} &= \pi s_{i+2}  T_{i+2}^{-1}=\pi T_{i+3}^{-1} s_{i+2}
,\quad \; i=1,2,3,4\, ,\\
s_{i+1} s_i  &=  \pi^{-1}   s_{i-1}T_{i}=  \pi^{-1} T_{i-1}  s_{i-1}
,\quad \; i=1,2,3,4 \, .
\end{split}
\label{nextsi}
\end{equation}
Furthermore,  we can reduce a product of three $s_i$'s to one $s_i$ sandwiched
between $\pi$'s and the shift operators:
\[
\begin{split}
s_i s_{i+2} s_{i+3}&= \pi^2 s_{i+1} T_{i+1}^{-1}  T_{i+4}^{-1}\\
s_i s_{i+1} s_{i+3}&= \pi^2 s_i T_i^{-1} T_{i+2}^{-1} 
\end{split}
\]
For completeness we also comment
an a product of three $s_i$ that has repeated indices and is
related to commuting pairs of $s_i$ as in
\[s_i s_{i+1}  s_i =\pi s_{i+2} s_{i+4} T_{i+2}^{-1} \, .\\
\]
For example $s_2 s_3 s_2=\pi s_4 s_2 T_4^{-1}$. Note also
that the products of B\"acklund transformations that are 
not next-to-nearest neighbors, namely $s_1 s_3$ and $s_2 s_4 $, 
are related to each other through:
\[s_1 s_3 =\pi^2 s_2 s_4 T_2^{-1} T_4^{-1}\, .\]

\subsection{Special seed solutions of PV equation}
\label{seeds}

The seed solutions    in terms of $q,p$  and their $\alpha_i$ 
parameters are written as  \cite{AGLZ}:
\begin{align}
(q,p)&=(z/2,z/2)_{\mathsf{a}},  \qquad ( \alpha_1, \alpha_2, \alpha_3, \alpha_4)= (\mathsf{a}, 1-\mathsf{a}, \mathsf{a}, 1-\mathsf{a})\, ,
\label{solution2} \\
(q,p)&=(z,0)_{\mathsf{a}}, \qquad \quad \quad   (\alpha_1, \alpha_2, \alpha_3, \alpha_4) =
(\mathsf{a}, 0,0, 2-\mathsf{a}) \, ,  %
\label{solution1}
\end{align}
where we found it instructive to include  an explicit dependence on the 
parameter $\mathsf{a}$ in expression for solutions: $(q,p)_{\mathsf{a}}$. 
In this notation it is easy to describe that  in this basis
$(T_1T_3)^{n}$ increases the parameter 
$\mathsf{a}$ : $\mathsf{a} \to \mathsf{a}+2 n$ but leaves $q=p=z/2$ 
of definition \eqref{solution2} unchanged:
\begin{equation} (T_1T_3)^{n} (z/2,z/2)_{\mathsf{a}}=
(z/2,z/2)_{\mathsf{a}+2n}\, .
\label{T1T3a}
\end{equation}
Similarly $(T_2T_4)^{n}$ decreases parameter  the $\mathsf{a}$ : $\mathsf{a}
\to \mathsf{a} - 2 n$ :
\begin{equation}  (T_2 T_4)^{n} (z/2,z/2)_{\mathsf{a}}=
(z/2,z/2)_{\mathsf{a}-2n} \, .
\label{T2T4a}
\end{equation}
These seed solutions \eqref{solution2} and \eqref{solution1} 
correspond to the $N=4$ dressing chain 
first-order polynomial solutions of \cite{AGLZ}
\begin{align}
&j_i= \frac{z}{4} (1,1,1,1)& 
&( \alpha_1, \alpha_2, \alpha_3, \alpha_4)= 
(\mathsf{a},1-\mathsf{a},\mathsf{a}, 1-\mathsf{a}) &  
\label{jsolution2} \\
& j_i= \frac{z}{2} (1,1,-1,1) & 
&  (\alpha_1, \alpha_2, \alpha_3, \alpha_4) =
(\mathsf{a},0,0,2-\mathsf{a}) &
\label{jsolution1} 
\end{align}
through an association
\[ q=j_1+j_2, \quad p=j_2+j_3 \]
with $j_i$ that satisfy the condition $\sum_{i=1}^4 j_i=z$.

The $\pi $ automorphism transforms $j_i, \alpha_i$ as follows
$(j_1,j_2,j_3,j_4 )\to (j_4 , j_1,j_2,j_3)$ and $ (\alpha_1, \alpha_2,
\alpha_3, \alpha_4) \to
 (\alpha_4, \alpha_1, \alpha_2, \alpha_3)$, 
and leads to the remaining  $\pi$ variants of \eqref{solution1}
\begin{align}
q&=z, \; p=z, \;(\mathsf{a}, 2-\mathsf{a},0,0)\, , %
\label{wata2}\\
q&=0, \;  p=z,\;  (0,\mathsf{a}, 2-\mathsf{a},0)  \label{wata5} \, ,\\
q&=0,\; p=0, \; (0,0, \mathsf{a},2-\mathsf{a})  \label{wata6} \, ,
\end{align}
that correspond to $j_i= (z/2) (1,1,1,-1), (z/2) (-1,1,1,1)$ and 
$(z/2) (1,-1,1,1)$, respectively.  One recognizes in the above 
solutions a  complete set of algebraic solutions found by 
Watanabe \cite{watanabe}.

As described in \cite{AGLZ} the class of solutions obtained by 
translations from \eqref{jsolution2} with all $j_i$ in `` up'' positions
gives rise to Umemura polynomials versus
solutions obtained by translations from \eqref{jsolution1} with one $j_i$
in `` down'' position, which are
special functions.
We find that all degeneracy will take place for solutions originating 
 from an initial configuration  with one $j_i$
`` down''  only, meaning it occurs only for 
solutions generated by symmetry transformations
from seed solutions having one $j_i$ `` down''.

\subsection{Classification of rational solutions of PV equation}
\label{classification}

The construction established in \cite{AGLZ} to produce the rational
solutions of PV  equation and explain their parameters falls into three classes.

For a class of rational solutions with their parameters : 
\begin{equation}
{ \alpha} = \frac12 (A)^2, \;\,  { \beta} =- \frac12
(A+n)^2 , \;\, { \gamma} =\epsilon m
\label{Ia}
\end{equation}
 and
\begin{equation}
 { \alpha} = \frac12 (B+n)^2, \;\,  { \beta} =- \frac12
(B)^2, \,\;{ \gamma} =\epsilon m
\label{Ib}
\end{equation}
where $A,B$ arbitrary parameters and $m+n$ is even, 
the corresponding solutions  can be  obtained by  
a successive operation by shift operators $T_i$ 
on polynomial seed solution \eqref{solution2}:
\begin{equation}
T_1^{n_1}T_2^{n_2}T_3^{n_3}T_4^{n_4}  
\left(q=\frac{z}{2},p=\frac{z}{2}\right)_{\mathsf{a}}, 
\qquad n_i \in  \mathbb{Z}\, ,
\label{T1234}
\end{equation}
From relations \eqref{Tialpha} we find that the configuration 
\eqref{T1234} possesses  parameters
\begin{equation}
{\alpha}_1 = \mathsf{a} +2 n_1-2 n_2,\, 
{\alpha}_3= \mathsf{a} +2 n_3-2 n_4,\;
{\alpha}_2= 1-\mathsf{a}+2 n_2-2 n_3,\,
{\alpha}_4= 1-\mathsf{a}+2 n_4-2 n_1\, ,
\label{pre-baralphas1}
\end{equation}
that reproduce parameters in \eqref{Ia}
with $A=\mathsf{a}/2+n_3-n_4$, $n=n_1+n_4-n_2-n_3$  and parameters in \eqref{Ib}
with $B=\mathsf{a}/2+n_1-n_2$, $n=n_2+n_3-n_1-n_4$ and
with $m=n_1+n_2-n_3-n_4$ in both cases.

These solutions are regular, meaning that they can be expanded in 
positive power series in $z$. %

If there is a rational solution with 
a pole singularity in $p$ it  can be removed by 
$s_1$  B\"acklund transformation (a pole singularity in $q$
can be removed by $s_2$  B\"acklund transformation).
Then the resulting regular configuration can be obtained
through action by the shift operators $T_i$  as in \eqref{T1234}
on seed solution \eqref{solution2} with the parameters
\begin{equation} { \alpha}= \frac12  \left( \frac{1}{2}+m\right)^2, \;
{ \beta}=- \frac12 \left( \frac{1}{2}+n \right)^2,  \;
{ \gamma}=(-A +n+m) \epsilon \, ,
\label{II}
\end{equation}
where $A$ is arbitrary and $n,m$ are integers.

Another class of rational solutions can 
obtained  from the seed solutions \eqref{solution1} 
by  successive operation with shift operators $T_i$
as in 
\begin{equation}
T_1^{n_1} T_4^{n_4} T_2^{-n_2} \left(q=z,p=0\right)_{\mathsf{a}}, 
 \qquad     n_2,n_4 \in  \mathbb{Z}_{+},
\; n_1  \in  \mathbb{Z}\, ,
\label{T1234restsol1}
\end{equation}
with final parameters given by 
\begin{equation} %
{\alpha}_1 = \mathsf{a} +2 n_1+2 n_2,\;{\alpha}_2= -2 n_2,\;
{\alpha}_3= -2 n_4,\; { \alpha}_4= 2-\mathsf{a} +2
n_4-2 n_1\, ,
\label{transalphas}
\end{equation}
Setting $n=\pm n_4 ,m=n_4+2n_2  \in \mathbb{Z}_{+}$ and with  $n+m$ 
being an even integer we conclude from   \eqref{transalphas}
that the resulting rational solutions are parameterized as follows: 
\begin{equation}
{ \alpha}= \frac18 \alpha_3^2 = \frac12  \left( n\right)^2, \;\;
{ \beta}=-\frac18 \alpha_1^2= - \frac12 \left(\epsilon { \gamma} +1 + m \right)^2,  \;\;
\label{IIIa}
\end{equation}
for the case of the seed solution \eqref{solution1}. Acting with
additional B\"acklund transformations 
$s_i,1,2,3$ on the parameters \eqref{transalphas} does not
change the expressions in \eqref{IIIa}, Acting with $s_4$ yields the
result given below in equation \eqref{IIIb}. Thus construction based solely on
translation operators captures the salient features of parameters
characterizing rational solutions as described in \cite{kitaev}.

For the remaining seed solutions $\left(q,p \right)_{\mathsf{a}}$ given in equations
\eqref{wata2}, \eqref{wata5},\eqref{wata6} the rational solutions 
\begin{equation}
T_i^{n_i} T_j^{n_j} T_k^{-n_k} \left(q,p \right)_{\mathsf{a}}, 
 \qquad     n_j,n_k \in  \mathbb{Z}_{+},
\; n_i  \in  \mathbb{Z}\, ,
\label{T1234restricted}
\end{equation}
are obtained for distinct $i,j,k$ such that $(i,j,k) =(2,1,3)$ for
\eqref{wata2}, $(i,j,k) =(3,2,4)$ for
\eqref{wata5} and $(i,j,k) =(4,3,1)$ for
\eqref{wata6}. The action by translation operator with $T_i^{n_i}$ with $n_i \in
\mathbb{Z}$ is special in that it only results in shifting the parameter $\mathsf{a}$ without
changing the form of solution $ \left(q,p \right)_{\mathsf{a}}$.
For
solutions generated from the seed solution \eqref{wata6} one obtains
\begin{equation}
{ \alpha}= \frac18 \alpha_3^2=\frac12  \left( -\epsilon { \gamma}+1+m\right)^2, \;\;
{ \beta}=-\frac18 \alpha_1^2=- \frac12 \left(  n \right)^2,  \;\;
\label{IIIb}
\end{equation}
In both cases of \eqref{IIIa} and \eqref{IIIa}
$n ,m  \in \mathbb{Z}_{+}$ and with  
$n+m$ being an even integer.

Note that in equations \eqref{T1234restricted} and equation
\eqref{T1234restsol1} only the shift operators 
that do not cause divergencies when acting on an appropriate seed 
solution appear while the other shift operators are  excluded. We
refer to this mechanism as exclusion rule. It is a main cause of degeneracy.

We conclude that the well-known  fundamental results
 on classification of rational solutions of Painlev\'e V equation
first  presented in \cite{kitaev}  
are here obtained by acting with translation operators \eqref{T1234}
on the first-order polynomial solutions \eqref{solution2} and \eqref{solution1}.
In the latter case we encounter 
restrictions on type of translations that are well-defined
and the allowed values of their powers $n_i$ 
as indicated in equation \eqref{T1234restricted}.

\subsection{Shift operators  acting on the solution
$(q=z/2,p=z/2)_{\mathsf{a}}$ }
\label{subsection:item1&2}

We have seen above in relation \eqref{T1234restsol1}  how   rational solutions were constructed 
out of $(q=z/2,p=z/2)_{\mathsf{a}}$ based on action of translation operators. Main issue to
consider is whether there are degeneracies associated with such
construction.

\begin{theor} There is no degeneracy among solutions generated 
by B\"acklund transformations out of the seed solution 
$(q=z/2,p=z/2)_{\mathsf{a}}$ from equation \eqref{solution2}.

We here study solutions generated from 
the seed solution \eqref{solution2} 
by action of $ T_1^{m_1}T_2^{m_2}$ $T_3^{m_3}T_4^{m_4} (q=z/2,p=z/2)_{\mathsf{a}}$ 
or by actions of  $s_{i_1} \cdots s_{i_r} 
T_1^{m_1}T_2^{m_2}T_3^{m_3}T_4^{m_4} (q=z/2,p=z/2)_{\mathsf{a}}$, 
where we include B\"acklund transformations $s_i, i=1,2,3,4$ and
their products.
We recognize that all the operators, translation operators, B\"acklund
transformations and their products are 
all invertible in this framework and therefore we do
not encounter any degeneracy. 
\end{theor}

Let us illustrate this on a following example:
\begin{exmp}
\label{example:s4a0}
We will act with the  B\"acklund transformation $s_4$ on
the seed solution $(q=z/2,p=z/2)_{\mathsf{a}=0}$. Since $\mathsf{a}=0$
then $\alpha_i=(0,1,0,1)$.
We find $s_4 (0,1,0,1)=(1,1,1,-1)$. Notice that $s_3(0,1,0,1)=(0,1,0,1)$
while $\pi s_3(0,1,0,1)=(1,0,1,0)$ and one more action by $s_3$ yields
$s_3 \pi s_3(0,1,0,1)=(1,1,-1,1)$. Thus $(\pi s_3)^2 (0,1,0,1)=(1,1,1,-1)$
and $(\pi s_3)^2 $ gives the same $\alpha_i$ as $s_4$.
We are going to show that they also produce the identical solution
acting on $(q=z/2,p=z/2)_{\mathsf{a}=0}$ in agreement with the
expectation of no degeneracy for solutions in this sector.
Indeed we easily find that $s_4$ and $(\pi s_3)^2 $ both yield
\[
s_4 (q=\frac{z}{2},p=\frac{z}{2})_{\mathsf{a}=0} = (\pi s_3)^2 (q=\frac{z}{2},p=\frac{z}{2})_{\mathsf{a}=0} 
= (q=\frac{z}{2}+\frac{2}{z},p=\frac{z}{2})_{\alpha_i=(1,1,1,-1)} 
\]
One should note that it holds quite generally that
$s_4= T_1 \pi (\pi s_3)^2 $ however $T_1\pi$ acts as an identity on
$(1,1,1,-1)$. Although  $s_4$ is not equal to
$(\pi s_3)^2 $ they produce identical solutions since the $\alpha_i$'s
of the transformed functions coincide.

The above expression is not regular and accordingly one consults
the case described in equation \eqref{II}, which is in agreement with the above
relations for the values of
$\alpha,\beta, \gamma$  obtained by inserting $A=-1$ and $n=0=m$.
\end{exmp}

As we will  now show it turns out that it is instructive 
to consider consequences 
of imposing conditions 
\begin{equation} s_i T_1^{n_1}T_2^{n_2}
T_3^{n_3}T_4^{n_4}  (\mathsf{a}, 1-\mathsf{a}, \mathsf{a}, 1-\mathsf{a})
= T_1^{m_1}T_2^{m_2}T_3^{m_3}T_4^{m_4}
(\mathsf{b}, 1-\mathsf{b}, \mathsf{b}, 1-\mathsf{b})
\label{siTTa}
\end{equation}
for solutions derived from
\eqref{solution2}. 
Inspecting the above relation
we find that it will have  solutions only for odd 
$\mathsf{a} =2 k+1, \mathsf{b}=2p+1$ for $s_i,
i=2,4$ and for even $\mathsf{a} =2 k, \mathsf{b}=2p$ for $s_i,
i=1,3$  for arbitrary integers $k,p$.

For B\"acklund transformations $s_i$ with $i$ even we can show   that:
\begin{equation}
s_i   \left(q=\frac{z}{2},p=\frac{z}{2}\right)_{\mathsf{a}}
=  (T_i T_{i+1}^{-1})^{(\mathsf{a}-1)/2} 
 \left(q=\frac{z}{2},p=\frac{z}{2}\right)_{\mathsf{a}}, \quad i=2,4
\label{degetestieven}
\end{equation}
for $\mathsf{a}=(2k+1)$ for which the power $(\mathsf{a}-1)/2$ is an
integer.

For B\"acklund transformations $s_i$ with $i$ odd it holds that:
\begin{equation}
s_i   \left(q=\frac{z}{2},p=\frac{z}{2}\right)_{\mathsf{a}}
=  (T_i^{-1} T_{i+1})^{\mathsf{a}/2} 
 \left(q=\frac{z}{2},p=\frac{z}{2}\right)_{\mathsf{a}}, \quad i=1,3
\label{degetestiodd}
\end{equation}
for $\mathsf{a}=2k$ for which the power $\mathsf{a}/2$ is an
integer.

The proof of relations \eqref{degetestieven} and \eqref{degetestiodd}
follows easily from applications of identities \eqref{lemmasi} and
\eqref{T1T3a} using the fact that due to commutation relations \eqref{lemmasi} one can
commute $s_i$ around a product of $T_{j_1} T_{j_2} \ldots$ till
finally ending up with $s_i$ acting on 
$\left(q=\frac{z}{2},p=\frac{z}{2}\right)_{\mathsf{a}=0}$ or 
$\left(q=\frac{z}{2},p=\frac{z}{2}\right)_{\mathsf{a}=1}$ 
where it can be replaced by identity  for $i$ odd or even, respectively.

For illustration we set $i=4$ and $\mathsf{a}=2k+1$, the proof 
extends easily to other values of $i$.
For $k=0$ or $\mathsf{a}=1$, we have $\alpha_i=(1,0,1,0)$ and
$s_4$  acts as identity 
 on $\left(q=\frac{z}{2},p=\frac{z}{2}\right)_{\mathsf{a}=1}$. 
 The proof of relation \eqref{degetestiodd} goes then 
 as follows. 
 \begin{equation}\begin{split}
&  s_4   \left(q=\frac{z}{2},p=\frac{z}{2}\right)_{\mathsf{a}=2k+1}
 = s_4 (T_1T_3)^{k}
 \left(q=\frac{z}{2},p=\frac{z}{2}\right)_{\mathsf{a}=1}=
 T_4^{k} T_3^{k} s_4
 \left(q=\frac{z}{2},p=\frac{z}{2}\right)_{\mathsf{a}=1}\\
 &=T_2^{k} T_1^{k} (T_1 T_3)^{-k} 
 \left(q=\frac{z}{2},p=\frac{z}{2}\right)_{\mathsf{a}=2k+1}=
(T_4 T_1^{-1})^{k} 
 \left(q=\frac{z}{2},p=\frac{z}{2}\right)_{\mathsf{a}=2k+1}\, ,
 \label{degetestinduction}
 \end{split}
 \end{equation}
which completes the proof.
Alternatively we could have just observed that since both sides of  
relations \eqref{degetestieven} and \eqref{degetestiodd}
have the same effect on the parameters $\alpha_i$ it follows
from uniqueness of solutions that these relations are correct.

Applying the same technique as in equation \eqref{degetestinduction}
but starting with  $s_4   \left(q=\frac{z}{2},p=\frac{z}{2}\right)_{\mathsf{a}=2k}$
and utilizing Example \ref{example:s4a0} we obtain
\[ s_4   \left(q=\frac{z}{2},p=\frac{z}{2}\right)_{\mathsf{a}=2k}=
 (\pi s_3)^2 (T_4 T_3^{-1})^{k} 
 (q=\frac{z}{2},p=\frac{z}{2})_{\mathsf{a}=2k} \, ,
\]
that generalizes \eqref{degetestiodd}  for an arbitrary even
$\mathsf{a}$.

It also follows that relations \eqref{degetestieven} and \eqref{degetestiodd}
are consistent with nilpotency of $s_i$  due to commutation relations
\eqref{lemmasi} as we now show for the particular case of $s_1$:
\[
\begin{split}
s_1^2  \left(q=\frac{z}{2},p=\frac{z}{2}\right)_{\mathsf{a}}&=
s_1 T_1^{-\mathsf{a}/2} T_{2}^{\mathsf{a}/2}
\left(q=\frac{z}{2},p=\frac{z}{2}\right)_{\mathsf{a}}=
T_2^{-\mathsf{a}/2} T_{1}^{\mathsf{a}/2} s_1
\left(q=\frac{z}{2},p=\frac{z}{2}\right)_{\mathsf{a}}\\
&=T_2^{-\mathsf{a}/2} T_{1}^{\mathsf{a}/2}T_1^{-\mathsf{a}/2} T_{2}^{\mathsf{a}/2}
\left(q=\frac{z}{2},p=\frac{z}{2}\right)_{\mathsf{a}}
= \left(q=\frac{z}{2},p=\frac{z}{2}\right)_{\mathsf{a}}\, .
\end{split}
\]
The other cases of $s_i, i=2,3,4$ follow {\it mutatis mutandis}.

These considerations extend to the products of B\"acklund transformations as 
easily deduced from relations \eqref{nextsi} and the ones shown below 
relations \eqref{nextsi}.  
\begin{exmp} 
For   the relation with $s_1 s_{2}$:  
\begin{equation} 
	s_1 s_2  T_1^{n_1}T_2^{n_2}
	T_3^{n_3}T_4^{n_4}  (\mathsf{a}, 1-\mathsf{a}, \mathsf{a}, 1-\mathsf{a})
	= T_1^{m_1}T_2^{m_2}T_3^{m_3}T_4^{m_4}
	(\mathsf{b}, 1-\mathsf{b}, \mathsf{b}, 1-\mathsf{b}) \, ,
	\label{s1s2TTa}
\end{equation}
chosen as a special example of the product of the type
$s_i s_{i+1} $, 
we find that for the relation \eqref{s1s2TTa}
to hold we must have an even $\mathsf{a}$ and an odd $\mathsf{b}$.
Since we can commute $s_1s_2$ around the translation operators, according to equation \eqref{lemmasi}, we focus on
\[
\begin{split}
s_1 s_2 (q,p)_{\mathsf{a}=2n }&=s_1 s_2 (T_1T_3)^n (q,p)_{\mathsf{a}=0}=
(T_2T_1)^n s_1 s_2  (q,p)_{\mathsf{a}=0}\\
&=(T_2T_1)^n T_1^{-1}   (q,p)_{\mathsf{b}=1}
=(T_2T_1)^n T_1^{-1} (T_2T_4)^n  (q,p)_{\mathsf{b}=2n+1}\\
&=T_2^{2n} T_1^{n-1} T_4^n (q,p)_{\mathsf{b}=2n+1} \,,
\end{split}
\]
where we used that 
\[
s_1 s_2 (0,1,0,1) =(-1,0,1,2)= T_1^{-1} (1,0,1,0) \, .
\]
\end{exmp}
We now consider the case of a product of $s_i$ that are not
next-to-neighbors.
\begin{exmp}
\label{example:s2s4}
We  consider $s_2 s_4$
for which it holds that the relation 
\begin{equation} 
	s_2 s_4  T_1^{n_1}T_2^{n_2}
	T_3^{n_3}T_4^{n_4}  (\mathsf{a}, 1-\mathsf{a}, \mathsf{a}, 1-\mathsf{a})
	= T_1^{m_1}T_2^{m_2}T_3^{m_3}T_4^{m_4}
	(\mathsf{b}, 1-\mathsf{b}, \mathsf{b}, 1-\mathsf{b}) \, ,
	\label{s2s4TTa}
\end{equation}
requires that $\mathsf{a}+\mathsf{b}-2=2(n_{2i}-n_{2i-1} +m_{4i}-m_{4i-1})$ for $i=1,2$ and that
$n_4-n_2=m_1-m_3$ and $n_1 -n_3 = m_4 -m_2$. For $n_i=0$ we find that $m_4=m_2, m_1=m_3$ and it follows
that
\[
s_2 s_4 (q,p)_{\mathsf{a}=2n }=
(T_2T_4)^{m_2} (T_1 T_3)^{m_1} (q,p)_{\mathsf{b}} \, ,
\]
with $\mathsf{a}+\mathsf{b}-2=2 (m_2-m_1)$. We calculate easily  that
$ s_2 s_4 (1,0,1,0)=(1,0,1,0)$
and thus $s_2 s_4$ is an identity on $\mathsf{a}=1$
consistent with $\mathsf{a}+\mathsf{b}-2=0$ for $\mathsf{b}=1$.
For $\mathsf{a}=0$ one finds
\[ s_2 s_4 (0,1,0,1)=(2,-1,2,-1) \, , \]  
which  amounts to mapping $\mathsf{a}=0$
to $\mathsf{a}=2$. This is consistent with relation
\[	s_2 s_4  (q,p)_{\mathsf{a}=0} =
	(T_2T_4)^{m_2} (T_1 T_3)^{m_1} (q,p)_{\mathsf{b}=0}	\, ,
\]
for $m_1-m_2=1$.
Using equations \eqref{T1T3a} and \eqref{T2T4a}
we can generalize these relations to arbitrary values of an integer $\mathsf{a}$ with  $s_2s_4$ being represented by products of 
the translation operators.
\end{exmp}
Let us now consider an arbitrary  product of B\"acklund transformations 
acting on the seed solution $ (q,p)_{\mathsf{a}}$:
$s_{i_1} s_{i_2} s_{l_3} \cdots s_{i_k}  (q,p)_{\mathsf{a}}$.
We can keep replacing each of the  products $s_i s_{i \pm 1}$ by a single $s_i$ multiplied by an  automorphism $\pi$ and an appropriate  translation operator according
to relations \eqref{nextsi}. In such way we end up with a single  B\"acklund transformation $s_k$ multiplying translations operators
if  the final   product we encounter is  $s_i s_{i+1} $ or $s_{i-1} s_i$.
If $\mathsf{a}$ is odd and $j$ even or vice versa we are then able to 
eventually replace all the $s_j$ transformations appearing in such process
completely by a product of translation operators and 
the case is  governed by relations
\eqref{Ia} or \eqref{Ib} resulting in Umemura polynomials 
for the positive powers of translation operators. 
If both $\mathsf{a}$ and $k$ is odd or even we end up with one single $s_k$ and product of translation operators and the
case is governed by relation \eqref{II}. If the final product of B\"acklund transformations
is  of the type $s_i s_{i+2} $ we  end up with only translation operators 
as illustrated by example \ref{example:s2s4}. If the  powers of 
translation operators are positive the result produces
Umemura polynomials.
These arguments explain why  creating rational solutions from
the seed solution $(q=z/2,p=z/2)_{\mathsf{a}} $ is achieved
by employing just translation operators or a single B\"acklund transformation
multiplied a product of translation operators.

\section{Derivation of conditions for degeneracy}
\label{vinculos}

 As will be described in this section the
cause of degeneracy is due to divergence 
one possibly  encounters in an attempt to equate two different
rational solutions, with identical $\alpha_i$'s, if one is to
violate the 
exclusion rule that needs to be upheld 
when generating rational solutions from the seed
solution \eqref{solution1}. %

Consider two standard quantities associated with an orbit. One
will describe the parameter of the orbit point
\begin{equation}
{\overline Y}_{m, \mathsf{b}} =  T_1^{m_1}T_2^{m_2}T_3^{m_3}T_4^{m_4}  
( \alpha_1, \alpha_2, \alpha_3, \alpha_4) 
\label{barYmi}
\end{equation}
and another describes a conventional solution given in \eqref{Ymdef}
with $(q=z,p=0)_{\mathsf{b}}$ being  the seed solution \eqref{solution1} with
its parameters given by 
$   (\alpha_1, \alpha_2, \alpha_3, \alpha_4) =
(\mathsf{b}, 0,0, 2-\mathsf{b})$.
Correspondingly integers $m_i$ will only take appropriate allowed values for which the 
quantity is finite. For the seed solution \eqref{solution1}
with $(\mathsf{b}, 0,0, 2-\mathsf{b})$ the allowed expression
is given by the following equation 
\begin{equation}
{\overline  Y}_m =T_2^{-m_2} T_4^{m_4}\,\left(\mathsf{b},0,0,2-\mathsf{b}\right) 
= \left( \mathsf{b}+2 m_2, -2 m_2, -2 m_4,
2- \mathsf{b}+2 m_4 \right), \;\; m_2,m_4 \in \mathbb{Z}_{+}\,.
\label{TTb}
\end{equation}
in agreement with equation \eqref{transalphas}.

We now consider more general expressions that also
include the products of $s_i$
transformations.
Schematically, we write these finite quantities as $X_{M, \mathsf{a}}$
given in  \eqref{XMdef} 
and we can represent $M$ from equation \eqref{XMdef} as a 
product of  B\"acklund transformations $s _i s_j s_k$ as shown below.
\begin{equation}
X_{n_i,\{i,j,k\} , \mathsf{a}} =    
s _i s_j s_k  T_1^{n_1}T_2^{n_2}T_3^{n_3}T_4^{n_4}  
(q,p)_{\mathsf{a}}
\label{Xni}
\end{equation}
In analogy to definition in \eqref{barYmi} we also introduce
a corresponding expression for the parameters of solutions
denoted as 
\begin{equation}
{\overline  X}_{n_i,\{i,j,k\} ,\mathsf{a}} =  s _i s_j s_k \ldots
T_1^{n_1}T_2^{n_2}T_3^{n_3}T_4^{n_4}  
( \alpha_1, \alpha_2, \alpha_3, \alpha_4) 
\label{barXni}
\end{equation}
with 
$   (\alpha_1, \alpha_2, \alpha_3, \alpha_4) =
(\mathsf{a}, 0,0, 2-\mathsf{a})$.

We impose 
\begin{equation}
{\overline  X}_{n_i,\{i,j,k\} ,\mathsf{a}} = {\overline  Y}_{m_i, \mathsf{b}}
\label{barXYni}
\end{equation}
to enforce that the parameters of both solutions  agree.

Note that an attempt to equate parameters \eqref{pre-baralphas1} (obtained
from expression \eqref{T1234} generated from solution
\eqref{solution2} $\left(q=\frac{z}{2},p=\frac{z}{2}\right)_{\mathsf{a}}$
to the  parameters ${\overline  Y}_m$ from equation \eqref{TTb} fails as it leads to
conflicting expressions (even and odd) for the value of $\mathsf{a}$.
Thus equating solutions derived from solutions derived from 
different types of seed solutions (ones with all positive $j_i$ and ones with one negative $j_i$) can not lead to degeneracy.

Degeneracy will occur when equating  
$X_{M , \mathsf{a}}$  and  $Y_{m, \mathsf{b}}$ will cause 
divergence when inverting some of the translation operators or
B\"acklund
transformations appearing on the left or right  hand side of an
anticipated equation
\begin{equation}
{X}_{n_i,\{i,j,k\} ,\mathsf{a}} \stackrel{?}{=} { Y}_{m_i, \mathsf{b}}
\label{XYmni}
\end{equation}
while the relation \eqref{barXYni} holds.

The appearance of divergence signals degeneracy 
as it prevents the left and right hand sides from being  equal despite 
 they sharing  the identical parameters due to the imposition
of condition \eqref{barXYni}.

The object $M$ that contains a B\"acklund 
transformation $s_i$ is to be determined from few general principles.
For products of the type $s_i s_{i+1} {\ldots} $ in definitions 
of $X_n$ 
we can reduce the product of two neighboring $s_i$'s  to the product
of $\pi$, one $s_i$ and the fundamental shift operator
as shown in relations \eqref{nextsi}.
One can show that terms of three $s_i$'s  will not lead to matching 
parameters, meaning that \eqref{barXYni}
will  not hold. For example the relation 
\begin{equation}
\begin{split}
&\pi^2 s_4 T_2^{-n_2} T_4^{n_4} \left( \mathsf{a}, 0,0,2-\mathsf{a}\right)=
\pi^2 \left( 2+2 (n_2+n_4), -2 n_2, 2-\mathsf{a},
-2+ \mathsf{a}-2 n_4 \right) \\&=
\left( 2- \mathsf{a},\mathsf{a}-2 -2n_4,2+2(n_2+n_4),- 2 n_2\right)\\
&=T_2^{-m_2} T_4^{m_4} \left( \mathsf{b}, 0,0,2-\mathsf{b}\right)
=\left( \mathsf{b}+2 m_2, -2 m_2, -2 m_4,
2- \mathsf{b}+2 m_4 \right)
\end{split}
\label{piinvcond}
\end{equation}
will require $0=1+n_2+n_4 +m_4$ that can not be solved for positive
integers and the same result will follow for expressions with arbitrary $s_i$ instead of $s_4$.
This observation excludes $\pi^2 s_i$ and $\pi^{-2} s_i$ and therefore $s_i s_{i+1} s_{i+2}$ and $s_i s_{i-1} s_{i-2}$ as candidates for $M$
that pairs with $Y_{m, \mathsf{b}}$ from definition  \eqref{Ymdef}.

After we established that the higher products of 
of adjacent $s_i$ reduce to $\pi^{\pm 1} s_i$ times shift operators
we proceed to determine for what $s_i$, the quantity  $X_M$ will remain
finite. Obviously, $s_2,s_3$ are not permitted in expressions of the type 
$ s_i   T_2^{-n_2} T_4^{n_4} (q=z,p=0)_{\mathsf{a}}$ since $s_i$ can be
commuted around the translation operators due to the 
commutation rules \eqref{lemmasi}. Thus we can only use
$s_1$ or $s_4$.
We are interested in a situation where equating 
$ s_i   T_2^{-n_2} T_4^{n_4} (q=z,p=0)_{\mathsf{a}}$ with 
$T_2^{-m_2} T_4^{m_4}\, (q=z,p=0)_{\mathsf{b}}$ for the same parameters
$\alpha_i$ will lead to divergencies that translate to degeneracy as
those two perfectly finite expressions can not be set equal although 
they satisfy the PV equation with the same parameters. 

To determine general  conditions for degeneracy let us equate 
expressions \eqref{Ymdef} and \eqref{XMdef} to each other:
\begin{equation}
X_M = M T_2^{-n_2} T_4^{n_4}\,
 (q=z,p=0)_{\mathsf{a}} = 
 T_2^{-m_2} T_4^{m_4}\, (q=z,p=0)_{\mathsf{b}}
\label{Mequation}
\end{equation}

After multiplication 
with $T_2^{m_2} T_4^{-m_4}$ on both sides of  equation \eqref{Mequation}
we get generally
\[
(q=z,p=0)_{\mathsf{b}}=  T_2^{m_2} T_4^{-m_4} M T_2^{-n_2} T_4^{n_4}\,
 (q=z,p=0)_{\mathsf{a}}= 
 M \,T_3^{c_3} \,T_2^{c_2}\, T_4^{c_4}  (q=z,p=0)_{\mathsf{a}}
 \]
obtained after commuting  $T_2^{m_2} T_4^{-m_4}$ around $M$ and ignoring
potential presence of $T_1$ on the right hand side since it only amounts
to shifting $\mathsf{a}$.
The conditions for degeneracy in this setting are 
\begin{equation}
c_3 \ne 0 , \;\, \text{or}\;\, c_2 >0\, , \;\, \text{or}\;\, c_4 <0 
\label{Mequationc}\end{equation}
since they correspond to presence of operators that will cause divergence
when acting on  $(q=z,p=0)_{\mathsf{a}}$. Next, we will investigate
values of $c_i, i=2,3,4$ for the 
possible candidates for $M$.

\subsection{Realization of solutions with degeneracy derived from  a seed solution
$(q=z,p=0)_{\mathsf{a}}$}
\label{subsection:realization}
As explained in \cite{AGLZ}  to avoid 
divergencies only action with  operators 
$T_1^{n_1} T_4^{n_4} T_2^{-n_2}$ with $n_2,n_4 \in \mathbb{Z}_{+}$ and
$n_1 \in \mathbb{Z}$ on solution \eqref{solution1} is well defined. 
Such operation generates parameters given above 
in equation \eqref{transalphas}.
The action of $T_1^{n_1} $ can be ignored as it only merely causes 
 shifting of $\mathsf{a}$ : 
$\mathsf{a} \to \mathsf{a}+2n_1$ 
without any change of $q=z,p=0$ \cite{AGLZ}. 
Also, as explained in \cite{AGLZ}, $s_2$ and $s_3$ are ill-defined on $
(q=z,p=0)_{\mathsf{a}}$ leaving only $s_1$ and $s_4$ as allowed
B\"acklund transformations.

We will consider solution ${Y}_{m_2,m_4, \mathsf{b}}$ from equation \eqref{Ymdef}
and attempt to equate it to \eqref{XMdef} after first imposing
condition \eqref{barXYni}.

For the B\"acklund transformations 
$s_1$ or $s_2$ in $M$ their action 
on $T_2^{-n_2} T_4^{n_4} (q=z,p=0)_{\mathsf{a}}$ lead 
to a finite result, but in order to, at the same time,  
generate an infinity on the
other side of equation \eqref{Mequation}
we need to turn $s_1,s_4$ into $s_2$ or $s_3$ 
so that they will create a divergence on the other side of the above 
equation. To accomplish that we substitute $s_i$ by
$\pi^{ \pm 1}  s_i$ that leads $s_1,s_4$ to become $s_2$ or $s_3$,
respectively, when commuting around $\pi^{ \pm 1}$.

Thus we are led to introduce two degenerate solutions
\begin{equation}
X_{+, n, \mathsf{a}}^{1}= \pi  s_1 T_2^{-n_2} T_4^{n_4}\,
 (q=z,p=0)_{\mathsf{a}},
\quad
X_{-, n, \mathsf{a}}^{4}= \pi^{-1}  s_4 T_2^{-n_2} T_4^{n_4}\,
 (q=z,p=0)_{\mathsf{a}},
\label{pipmsi}
\end{equation}
where we used $M= \pi s_1= s_2 \pi$ and $M=\pi^{-1} s_4=s_3 \pi^{-1}$
to end up with $s_2$ and $s_3$ producing infinity on the right hand
side of the proposed equation
\begin{equation}
X_{ \pm, n, \mathsf{a}}^{i}= {Y}_{m_2,m_4, \mathsf{b}}, \;\; i=2,4\,.
\label{proposedeq}
\end{equation}

There is another potential source of infinity connected with 
multiplication with $T_4^{-m_4}$ on both side of equation
\begin{equation}
X_{  n}^{1,2}=  s_1 s_2 T_2^{-n_2} T_4^{n_4}\,
 (q=z,p=0)_{\mathsf{a}} = 
 T_2^{-m_2} T_4^{m_4}\, (q=z,p=0)_{\mathsf{b}}
 \label{X12Y}
\end{equation}
as it  results in the term $T_4^{n_4-m_4}$ that for $n_4-m_4 <0$ causes 
divergence and corresponds to $ c_4 <0$ in equation \eqref{Mequationc}.
Note that $s_1 s_2$ commutes with $T_4$ according to relations
\eqref{lemmasi}.

Although the expression \eqref{X12Y} appears at first to be infinite due
to the presence of $s_2$, however since $  s_1 s_2 T_2^{-1}=T_3^{-1}s_1 s_2= 
s_3 \pi^3 s_1 s_2 s_1 s_2= s_3 \pi^3s_2 s_1$
the divergencies cancel.
Another way to see that  the expression on the left hand side of \eqref{X12Y}
is finite despite the
presence of $s_2$ is to notice that 
 moving $s_1 s_2$ around $T_2^{-n_2} T_4^{n_4}$
generates $T_3$ and divergencies from $T_3$ and $s_2$ cancel each
other as long as $n_2>0$. Multiplying with $T_4^{-m_4}$ on both sides causes a divergence
since $m_4>n_4$ in order for the $\alpha_i$ parameters agree. This causes 
 the expression  \eqref{X12Y} to be afflicted by degeneracy as solutions on the left and the right hand side of this 
 equations can not be set equal.

One could also consider along the same lines
\[ X_{  n}^{3,4}=  s_3 s_4 T_2^{-n_2} T_4^{n_4}\,
 (q=z,p=0)_{\mathsf{a}} = 
 T_2^{-m_2} T_4^{m_4}\, (q=z,p=0)_{\mathsf{b}}
\]
expecting a divergence caused by $T_2^{m_2}$ multiplication however
since $s_3 s_4= \pi s_1 T_1^{-1}$ this case has already been considered.

Another proposal is \[
X_{\pm,   n}= \pi^{\pm} T_2^{-n_2} T_4^{n_4}\,
 (q=z,p=0)_{\mathsf{a}} = 
 T_2^{-m_2} T_4^{m_4}\, (q=z,p=0)_{\mathsf{b}}
\]
since due to identities $\pi T_i= T_{i+1} \pi$ and $ T_i\pi^{-1}= T_{i+1} 
\pi^{-1}$,  multiplication of  $X_{+,   n}$ by $T_4$  will produce $T_3$
and  multiplication of  $X_{-,   n}$ by $T_2$  will also produce $T_3$.
Note however that we can rewrite e.g. $X_{+,   n}$ as $X_{+, n, \mathsf{a}}^{1}$ due to
\[
\begin{split}
&\pi T_2^{-n_2} T_4^{n_4}\,
 (q=z,p=0)_{\mathsf{a}}= \pi T_2^{-n_2}
 T_4^{n_4}\,T_1^{\mathsf{a}/2}  (q=z,p=0)_{\mathsf{a}=0}\\
 &
 = \pi T_2^{-n_2}
 T_4^{n_4}\,T_1^{\mathsf{a}/2} s_1 (q=z,p=0)_{\mathsf{a}=0}
 =\pi s_1 T_1^{-n_2} T_4^{n_4}\,T_2^{\mathsf{a}/2}  (q=z,p=0)_{\mathsf{a}=0}
 \end{split}
\]
with expression valid for even $\mathsf{a}<0$, 
which is in agreement with value of  $\mathsf{a}$
required for matching parameters of $X_{+,   n}$ with
$Y_{m_2,m_4}$.

These considerations point toward three configurations with  solutions
exhibiting degeneracy. They are listed below together with conditions
required for  the agreement of $\alpha_i$ parameters :
\begin{table}[h]
	\centering
	\ra{1.5}
	\begin{tabular}{ll}
		\toprule
		$X_n$ & conditions for $ \overline{X}_n = \overline{Y}_{ m_2, m_4, \mathsf{b} }$  \\
		\midrule
		$X_{+, n}^{1}= \pi  s_1 T_2^{-n_2} T_4^{n_4}(q=z,p=0)_{\mathsf{a}}$
		& $\mathsf{a}=2 (m_2-n_2)$, \; $\mathsf{b}= 2(1-m_2+n_2+n_4)$, \\
		& $m_4=n_2-m_2$ \\
		\addlinespace[.8em]
		$X_{-, n}^{4}= \pi^{-1}  s_4 T_2^{-n_2} T_4^{n_4}(q=z,p=0)_{\mathsf{a}}$
		& $\mathsf{a}=2(1+n_4-m_4)$,\;$\mathsf{b}= 2(m_4-n_2-n_4)$, \\
		& $m_2=n_4-m_4$ \\
		\addlinespace[.8em]
		$X_{n}^{1,2}=  s_1 s_2 T_2^{-n_2} T_4^{n_4} (q=z,p=0)_{\mathsf{a}}$
		& $\mathsf{a}=-2(n_2+m_2)$, \; $\mathsf{b}= 2(m_4-n_4)$, \\
		& $m_4=n_2+n_4$ \\
		\bottomrule
	\end{tabular}
	\caption{Conditions for $\overline{X}_n = \overline{Y}_{ m_2, m_4, \mathsf{b} }$  }
\label{tableXn}
\end{table}

Common for all these cases is that matching the $\alpha_i$ parameters
through relations   ${\overline  X}_{+, n}^{1} =
{\overline  Y}_{m}$, ${\overline  X}_{-, n}^{4} =
{\overline  Y}_{m}$ and ${\overline  X}_{n}^{1,2} =
{\overline  Y}_{m}$ requires that both $\mathsf{a}$ and  $\mathsf{b}$
are  even numbers.

We will now systematically derive conditions for degeneracy that
follow from equation \eqref{Mequation} as derived in \eqref{Mequationc}
for $ M= \pi  s_1, \pi^{-1}  s_4$ and $ s_1 s_2$.

For $ M= \pi  s_1$ we find after multiplication by $T_2^{m_2} T_4^{-m_4}$
on the left hand side of equation \eqref{proposedeq}:
\[ T_2^{m_2} T_4^{-m_4} \pi  s_1 T_2^{-n_2} T_4^{n_4}(q=z,p=0)_{\mathsf{a}}=
\pi  s_1 T_3^{-m_4}  T_2^{m_2 -n_2} T_4^{n_4}(q=z,p=0)_{\mathsf{a}}\]
Thus $m_4 \ne 0 $ or $m_2 >n_2$ are the conditions for the degeneracy, 
(corresponding to $c_3 \ne 0 $ and $c_2 >0$ in equation
\eqref{Mequationc}),
however the latter  condition is  not met since from
the first row of table \ref{tableXn} we see that $m_4=n_2-m_2 \ge 0$ and
thus the only valid remaining condition is $m_4 \ne 0 $.

For $ M= \pi^{-1}  s_4$ we find
\[ T_2^{m_2} T_4^{-m_4} \pi^{-1}  s_4 T_2^{-n_2} T_4^{n_4}=
\pi^{-1}  s_4  T_3^{m_2}  T_2^{-n_2} T_4^{-m_4+n_4}\]
The divergence will occur for either  $m_2 \ne 0 $ or $m_4 >n_4$, 
 however since from
the second row of table \ref{tableXn} we know that $m_2=n_4-m_4 \ge 0$
it is not possible to have $m_4 >n_4$ and we
are left with only one valid condition $m_2 \ne 0 $, or equivalently
$n_4 >m_4$.

Finally for $ M= s_1  s_2$ we find
\[ T_2^{m_2} T_4^{-m_4} s_1  s_2 T_2^{-n_2} T_4^{n_4}=
s_1  s_2   T_2^{m_2-n_2} T_4^{-m_4+n_4}\]
Infinities occur for $m_4>n_4$ or  $m_2>n_2$. From
the third row of table \ref{tableXn} we know that $m_4-n_4=n_2 \ge 0$.
Thus we need $n_2 >0$ or $m_2>n_2$ for degeneracy to occur. Note that
$ s_1  s_2 T_2^{-n_2} T_4^{n_4} (q,p)_{\mathsf{a}}$ is only well-defined for
$n_2 >0$, which also happens to be a condition for degeneracy.

We now show in Table \ref{tableXnpara}
the $\alpha_i$ parameters of the degenerated cases and the conditions that
integers $n_2,n_4,$ $m_2, m_4 \in \mathbb{Z}_{+}$ must satisfy for
the degeneracy to occur:

\begin{table}[h]
	\centering
	\ra{1.6}
	\begin{tabular}{lll}
		\toprule
		$X_n$ & $\alpha_i$ parameters of ${\overline  X}_n = {\overline  Y}_{m}$ & conditions for degeneracy \\
		\midrule
		$X_{+, n}^{1}$  &  $2 ( 1+n_2+n_4, -m_2, m_2-n_2, -n_4)$ &  
		\quad $n_2 >m_2 \ge 0$ \\
		$X_{-, n}^{4}$  & $2(-n_2, m_4-n_4,-m_4,1+n_2+n_4)$ & 
			\quad $n_4 >m_4\ge 0$ \\
		$X_{  n}^{1,2}$ & $2(n_2+m_2,-m_2,-n_2-n_4,1+n_4)$ & 
			\quad  $n_2>0$ \\
		\bottomrule
	\end{tabular}
	\caption{Conditions for degeneracy}
	\label{tableXnpara}
\end{table}

We see that  the parameters in the second column in  the third row
are  different from $\alpha_i^{(1)}$, $\alpha_i^{(4)}$, which both
have three negative components. The parameters in the first two rows will agree 
after acting with an automorphism $\pi$ and performing change of variables
given below in \eqref{nmchange}. We will show below that the first two rows
represent a two-fold degeneracy for each value of $\alpha_i$'s.

\subsection{A two-fold degeneracy of $X_{+, n}^{1}$ and $X_{-, n}^{4}$
solutions}
\label{subsection:x1x4}
Consider the parameter $\alpha_i^{(4)}$ of solution $X_{-, n}^{4}$ from the second row of
Table \ref{tableXnpara}
\begin{equation} \alpha_i^{(4)} = 2\left(-n_2,  m_4-n_4, -m_4, 1+n_2+n_4\right),  \; \; 
m_4<n_4,  \; \; 
n_2,n_4, m_4 \in
\mathbb{Z}_{+}\, ,
\label{alphai4}
\end{equation}
and perform the following change of variables 
\begin{equation} n_2 \to {\bar m}_2,\;  m_4 \to {\bar n}_4,\; 
n_4 \to {\bar n}_2 +{\bar n}_4 - {\bar m}_2, %
\label{nmchange}
\end{equation}
which  transforms
$\alpha_i^{(4)} $ to
\begin{equation}
\alpha_i^{(4)} = 2\left(-{\bar m}_2,  {\bar m}_2-{\bar n}_2 , -{\bar n}_4
, 1+{\bar n}_2 +{\bar n}_4\right),  \; \; 
{\bar n}_2 > {\bar m}_2 \, ,
\label{alpha4bar}
\end{equation}
which agrees with $\alpha_i^{(1)} $ (in a barred notation) after acting
with $\pi$ automorphism.
Also we observe that the relation $m_2=n_4-m_4 \ge 0$ from second row of
Table \ref{tableXnpara} transforms 
to ${\bar n}_2 +{\bar n}_4-{\bar m}_2-{\bar n}_4 \ge 0$ or
${\bar n}_2 \ge {\bar m}_2$.
Thus the degeneracy condition $n_4 > m_4$ translates to 
${\bar n}_2 +{\bar n}_4-{\bar m}_2 >{\bar n}_4$
or ${\bar n}_2> {\bar m}_2 \ge 0$, which is a correct degeneracy
condition for the first row of the 
Table \ref{tableXnpara} with $\alpha_i^{(1)}$. Thus we had established
equivalence of the first two rows of 
the Table \ref{tableXnpara}.

Further one establishes for 
the four solutions 
$X_{n}^{(1)}, Y^{(1)}_m, \pi X_n^{(4)}$ and 
$\pi Y_m^{(4)}$, that  share the same parameter 
$\alpha_i^{(1)}$ as in equation \eqref{alphapi1},
will satisfy  two equalities between $Y^{(1)}_m$ and  $\pi X_n^{(4)}$
and between $\pi Y_m^{(4)}$ and $X_{n}^{(1)}$ resulting in:
\begin{equation}
Y^{(1)}_m = \pi X_n^{(4)} \; \ne \; \pi Y_m^{(4)}=X_{n}^{(1)}
\label{two-fold}
\end{equation}
These relations  show that the 
the degeneracy is a two-fold degeneracy of the parameter
$\alpha_i^{(1)}$ from equation \eqref{alphapi1}.

\begin{exmp} For the case of $X_{-, n}^{(4)}$ choose 
	\[  m_4=0, \;n_4=1,\; n_2=1,\; \to \;\mathsf{a}=4, \mathsf{b}=-4,
	\]
	so that $n_4>m_4$ and $m_2 =n_4-m_4=1$, which satisfy
	conditions  for	degeneracy.
	
	The corresponding solutions are :
	\begin{equation} \pi^{-1} s_4 T_4^{1} T_2^{-1} (q= z,p=0)_{\mathsf{a}=4}
	\ne  T_2^{-1} T_4^{0} (q=z,p=0)_{\mathsf{b}=-4}
	\label{expiis4}
	\end{equation}
	with $\alpha_i=( -2,-2,0, 6)$ for both sides.
	
	We find for the left hand side of inequality \eqref{expiis4}:
	\[ 
	q= -\frac{2 z (-4z^2+z^4+8)}{(-2+z^2) (-8z^2+z^4+8)}, \quad
	p= \frac{2 z (-8 z^2+z^4+8)}{(z^2-4)(-4z^2+z^4+8)}
	\]
	while on the right hand side of  \eqref{expiis4} we find:
	\[ q=z, \; p = \frac{2 z}{-4-z^2}
	\]
	that also satisfies PV  equation with $\alpha_i^{(4)}=2 ( -1,-1,0, 3)$.
According to the transformation \eqref{nmchange} the above parameters
transform to 
\[ (n_2,n_4, m_2,m_4)=(1,1,1,0) \to
({\bar n}_2,{\bar n}_4, {\bar m}_2, {\bar m}_4)=(2,0,1,1)
\]
and inserting the barred quantities into 
$\alpha_i^{(1)} $ in equation \eqref{alphapi1} gives:
$2(3, -1,-1,0)$ which agrees with the above found $\alpha_i^{(4)}$ up
to a $\pi$ automorphism.

\end{exmp}
Consistent with equivalence between solutions  $X_{+, n}^{1}$
and $X_{+, n}^{4}$ 
we only need to consider detailed examples associated
with  solutions  $X_{+, n}^{1}$
to which we turn in the next subsection.

The next two subsections will deal with examples of degeneracy
of solutions $X_{+, n}^{1}$ and $X_n^{1,2}$.

\subsection{Examples of   degeneracy for
${\overline X}_{+, n}^{1}=\pi s_1  T_2^{-n_2} T_4^{n_4} (\mathsf{a},0,0,2-\mathsf{a})$ equal to ${\overline  Y}_m =T_2^{-m_2} T_4^{m_4}(\mathsf{b},0,0,2-\mathsf{b}) $
}
\label{subsection:s3s4T2T4}

Recall from Table \ref{tableXnpara} that $\alpha_i^{(1)} = {\overline X}_{+, n}^{1}={\overline  Y}_m=
2(n_2+m_2,-m_2,-n_2-n_4,1+n_4)$ %
and the condition for degeneracy is $n_2 >m_2 \ge 0$.

\begin{exmp}
Let us first take for simplicity a special  case of $n_4=0$ %
and $m_2=0$.
Then the solutions are
\[X_{+, n}^{1}  (q=z,p=0)_{\mathsf{a}}= 
\pi s_1 T_2^{-n_2}   (q=z,p=0)_{\mathsf{a}}\]
with $\mathsf{a}=-2$ and
\[ Y_m   (q=z,p=0)_{\mathsf{b}}=
T_2^{-n_2+1}  T_4   (q=z,p=0)_{\mathsf{b}}\]
with $ \mathsf{b}=4$.

Setting $n_2=1$ (such that $n_2> m_2=0$) we find
\begin{equation} X_n^{(1)}= \pi s_1 T_2^{-1}   (q=z,p=0)_{\mathsf{a}=-2}
=\left(q= z\frac{z^2+4}{z^2+2}, p=z \right) 
\label{s3s4T2inv}
\end{equation}
and 
\begin{equation}
Y_m= T_4   (q=z,p=0)_{\mathsf{b}=4}
= \left(q= z\frac{z^2-4}{z^2-2}, p=0 \right)
\label{T4b4}
\end{equation}
which both solvethe  PV equation with $\alpha_i=\left(4,0, -2 ,0\right)$.
\end{exmp}

Next we consider a more general case of $n_2=n_4=n+1$:
\begin{equation}
\begin{split}
{\overline X}_n^{(1)}&=\pi s_1  T_2^{-n-1} T_4^{n+1} (\mathsf{a},0,0,2-\mathsf{a})\\&=
\pi s_1  (\mathsf{a}+2(n+1),-2 (n+1),-2(n+1) ,2(n+1) +2-\mathsf{a})\\&=
 \pi (-2 (n+1)-\mathsf{a}, \mathsf{a}, -2 (n+1),4 (n+1) +2)\\&=
 (4 (n+1) +2,-2 (n+1)-\mathsf{a}, \mathsf{a},  -2(n+1))
\end{split}
\label{s3s4T2nT4n}
\end{equation} 
and compare with 
\begin{equation}
{\overline Y}_m= T_2^{-m} T_4^{m} (\mathsf{b},0,0,2-\mathsf{b})=
(\mathsf{b}+2m,-2 m,-2m ,2m +2-\mathsf{b})
\label{T2mT4m}
\end{equation} 
for   $m_2=m_4=m$.

Imposing  the condition: $ {\overline X}_n={\overline Y}_m$
is equivalent in components to:
\begin{align}
4 n +6&=\mathsf{b}+2m , \qquad %
\,-2n-2=2m +2-\mathsf{b}  ,\label{a.2} \\
\mathsf{a}&=-2 m, \qquad %
-2n -2 -\mathsf{a}=-2 m \, . \label{a.4} 
\end{align}
Summing the two equations \eqref{a.2} we get \begin{equation}
m=\frac{n+1}{2} \, ,
\label{b.0}
\end{equation}
which obviously is consistent with $m_4=n_2-m_2$ as generally
established above.
Plugging relation \eqref{b.0} back into equations  \eqref{a.2}-\eqref{a.4} we obtain:
\begin{equation} \mathsf{a} = -1-n , \quad
\mathsf{b} = 3n+5 \label{b.2}
\end{equation}
Inserting these values for $\mathsf{a}$ and $\mathsf{b}$ into
expressions \eqref{s3s4T2nT4n} for parameters of solutions we obtain
generally:
\begin{equation}
\alpha_i= (4 n+6, -n-1,-n-1,-2n-2), \quad n=2k+1,\; k=0,1,2\, .
\label{alphaclarkson}
\end{equation}

Let us calculate explicitly the case of $n=1$ which implies
$\mathsf{a}=-2, \mathsf{b}=8$, $m=1$ %
and $\alpha_i= (10, -2,-2,-4)$.
According to expressions \eqref{s3s4T2nT4n} and \eqref{T2mT4m}  we have two solutions
\begin{equation}
\begin{split}
X_{n=1}^{(1)}&=\pi s_1 T_2^{-2} T_4^{2} (q=z,p=0)_{\mathsf{a}=-2} \, ,\\
Y_{m=1} &= T_2^{-1} T_4^{1}(q=z,p=0)_{\mathsf{b}=8}\, .	
\end{split}\label{Xn1Ym1}
\end{equation}
We first calculate $Y_{m=1}$ starting with 
\begin{equation}
T_2^{-1}: q=z, p=0 \to q=z, p= \frac{2z}{\mathsf{b} -z^2}, \, \; \;
(2+\mathsf{b},-2,0,2-\mathsf{b}) \, .
\label{T2m1}
\end{equation}
Action of $T_4$ on $q,p$ gives generally
\begin{equation}
\begin{split}
T_4(q)&= z-p-(\alpha_1+\alpha_4)/(q+\alpha_4/(z-p)) \\
T_4 (p) &=
q+\alpha_4/(z-p)-(\alpha_1+\alpha_2+\alpha_4)/(p+(\alpha_1+\alpha_4)/(q+\alpha_4/(z-p)))
\label{T4pq}
\end{split}
\end{equation}
Acting on configuration \eqref{T2m1} with $T_4$ according to \eqref{T4pq}
we get
\begin{equation}
\begin{split}
q&=z\frac{(-\mathsf{b}+z^2+2)(z^4-2 z^2 \mathsf{b}+\mathsf{b}^2+2\mathsf{b})}{(-\mathsf{b}+z^2)(-2z^2
\mathsf{b}+z^4+4z^2-2\mathsf{b}+\mathsf{b}^2)}\\
p&=-2z \frac{(-2 z^2 \mathsf{b}+z^4+4 z^2-2\mathsf{b}+\mathsf{b}^2)}{(-\mathsf{b}+z^2+2)(-2z^2
\mathsf{b}+z^4-2\mathsf{b}+\mathsf{b}^2)}
\label{ppqq}
\end{split}
\end{equation}
which for $\mathsf{b}=8$ yields
\begin{equation}
q= \frac{z (z^2-6) (z^4-16 z^2+80)}{(z^2-8)(z^4-12 z^2+48)}, \;\;
p =  \frac{-2 z(z^4-12 z^2+48)}{ (z^2-6)(z-2)(z+2)(z^2-12)} \, ,
\label{solutions11}
\end{equation}
with $\alpha_i= (10, -2,-2,-4)$.

Next we calculate $X_{n=1}$ acting first
with $T_2^{-2}$ on  $q=z,\; p=0$. Recall from \cite{AGLZ}
that :
\[
T_2^{-n}: q=z, p=0 \to q_n=z, p_n=\frac{2n z R_{n-1} (\mathsf{a};z)}{R_{n}
(\mathsf{a};z) },  \, 
(2n+\mathsf{a},-2n,0,2-\mathsf{a})\, ,
\]
(see \cite{AGLZ} for definition of $R_{n}
(\mathsf{a};z)$).  The above equation yields for $n=2$ :
\begin{equation}
T_2^{-2}: q=z, p=0 \to q=z, p=\frac{4 z (\mathsf{a}-z^2)}{z^4 -2 \mathsf{a}
z^2+\mathsf{a}(\mathsf{a}+2) },  \, 
(4+\mathsf{a},-4,0,2-\mathsf{a})\, ,
\label{T2m2}
\end{equation}
Furthermore we find 
\[T_4^2 (4+\mathsf{a},-4,0,2-\mathsf{a})=
(4+\mathsf{a},-4,-4,6-\mathsf{a}) \, .
\]
Applying $T_4^2$, using expression \eqref{T4pq}, on the configuration in
equation \eqref{T2m2} we get a complicated a solution to Painlev\'e
equation for $\alpha_i=(4+\mathsf{a},-4,-4,6-\mathsf{a})=(2,-4,-4,8)$,
which for $\mathsf{a}=-2$ is equal to 
\begin{equation}
\begin{split}
q&=z\,\frac{(z^4+12\,z^2+48)\,(z^8+16\,z^6+96\,z^4+192\,z^2+192)}
{(z^8+24\,z^6+216\,z^4+768\,z^2+1152)\,(8\,z^2+24+z^4)}\, ,\\
p&=-4\,\frac{(z^6+6\,z^4+24\,z^2+48)\,(z^8+24\,z^6+216\,z^4+768\,z^2+1152)}
{z\,(z^6+12\,z^4+72\,z^2+192)\,(z^8+16\,z^6+96\,z^4+192\,z^2+192)} \, ,
\end{split}
\label{solsT2T4}
\end{equation}
which exactly, as expressions \eqref{solutions11},
solves the PV equation with $\alpha_i= (10, -2,-2,-4)$.

Applying then $\pi s_1$ transformation $(2,-4,-4,8)\to (10,-2,-2,-4)$ we are
being taken from solution \eqref{solsT2T4} to %
\begin{equation}
\begin{split}
 q&=
z\,\frac{(8\,z^2+24+z^4)\,(z^6+18\,z^4+144\,z^2+480)}{(z^4+12\,z^2+48)\,(z^6+12\,z^4+72\,z^2+192)}
\, , \\
p&=z\,\frac{(z^4+12\,z^2+48)\,(z^8+16\,z^6+96\,z^4+192\,z^2+192)}{(z^8+24\,z^6+216\,z^4+768\,z^2+1152)\,(8\,z^2+24+z^4)}
\, .
\label{solutions2}
\end{split}
\end{equation}
Acting further with $s_3$ we get $(10,-2,-2,-4) \to (10,-4,2,-6)$
in agreement with  $\alpha_i$ in \eqref{alphaclarkson}.
Acting  with $s_3$ on \eqref{solutions2} we get
\begin{equation}
\begin{split}
q &=
z\,\frac{(8\,z^2+24+z^4)\,(z^6+18\,z^4+144\,z^2+480)}{(z^4+12\,z^2+48)\,(z^6+12\,z^4+72\,z^2+192)}
\, ,\\
p &= -4\,\frac{(z^4+12\,z^2+48)}{(z\,(8\,z^2+24+z^4)} \, .
\label{solutions1}
\end{split}
\end{equation}
Acting  with $s_3$ on \eqref{solutions11} we get
\begin{equation}
q= \frac{z (z^2-6) (z^4-16 z^2+80)}{(z^2-8)(z^4-12 z^2+48)}, \;\;
 p =  \frac{(z^4-12 z^2+48)}{ z(z^2-6)} \, .
\label{solutions111}
\end{equation}
Expressions \eqref{solutions1}, \eqref{solutions111} with $\alpha_i=(10,-4,2,-6)$.
are the ones first found in \cite{CD}.

For $n=3$ it holds $\mathsf{a}=-4, \mathsf{b}=14$ and $m=2$, %
Generally, for $n=2k+1$ it holds $\mathsf{a}=-2(k+1), \mathsf{b}=8+6k$ and 
$m=k+1$. %

With  solutions shown above in \eqref{b.0}-\eqref{b.2} and labeled by
$k=0,1,2,{\ldots} $ with $\mathsf{b}=2-3 \mathsf{a}$,
$\mathsf{a}=-2(k+1)$
and $n=2k+1=-1-\mathsf{a}$ we can compare the two solutions constructed out of 
$(q=z,p=0)_{\mathsf{a}}$ and out  of $(q=z,p=0)_{\mathsf{b}=2-3 \mathsf{a}}$. 
Explicitly one solution is  :
\begin{equation}
\pi s_1 (T_2^{-1} T_4)^{(n+1)}  (q=z,p=0)_{\mathsf{a}}=
T_1^{n+1}  T_2^{-n-1}  \pi s_1 (q=z,p=0)_{\mathsf{a}}
\label{s3s4}
\end{equation}
where we used used relations  \eqref{lemmasi} 
to move $T_2,T_4$ around $\pi,s_1$. The other solution after
substituting \eqref{b.0} for $m$ is 
\begin{equation}
 (T_2^{-1} T_4)^{\frac{n+1}{2}}  (q=z,p=0)_{\mathsf{b}}
\label{T2T4n12}
\end{equation}
Although both solutions \eqref{s3s4} and \eqref{T2T4n12}
share the same parameters $\alpha_i =
(4 n+6 , -n -1, -n-1, -2(n+1))$ with conditions \eqref{b.0}-\eqref{b.2}
being satisfied however they can not be equal. If they would be equal
then we would be free to multiply both sides by $T_2^{n+1}$ 
leaving essentially a finite expression $\pi  s_1 (q=z,p=0)_{\mathsf{a}}$.
However this would result in $T_2^{n+1-m}$
on the right hand side acting on $(q=z,p=0)_{\mathsf{b}}$.
Since $m=(n+1)/2$ then the exponent 
of $T_2$ will be equal to a positive number $(n+1)/2 >0$ which must be excluded
from acting on $ (q=z,p=0)_{\mathsf{b}}$ as it is causing divergencies. This provides a simple and explicit proof for degeneracy of  solutions \eqref{s3s4} and \eqref{T2T4n12}.

\subsection{Example of degeneracy for  
${\overline X}_n^{1,2}=s_1 s_2 T_2^{-n_2} T_4^{n_4} (\mathsf{a},0,0,2-\mathsf{a}) $ equal to 
 ${\overline Y}_m =T_2^{-m_2} T_4^{m_4}(\mathsf{b},0,0,2-\mathsf{b}) $
}
\label{example:s1s2}
This case seems at first to be infinite due to presence of $s_s$ but
the term $s_2 T_2^{-1}=s_2s_2s_3 \pi^3 s_1= s_3 \pi^3 s_1$ does not lead to infinities
as $s_2$ is canceled out as long as $n_2>0$ (which happens to be precisely the
condition for degeneracy).

We consider the corresponding parameters given by:
\[
\begin{split}
{\overline X}_n^{1,2} (\mathsf{a},0,0,2-\mathsf{a}) =& (-a, a+2n_2,-2n_2-2n_4,2+2n_4),\\
{\overline Y}_m (\mathsf{b},0,0,2-\mathsf{b})&=
(\mathsf{b}+2m_2,-2m_2,-2m_4,2-\mathsf{b}+2m_4) \,.
\end{split}
\]
To set these two expressions equal requires:
\begin{equation}
m_4=n_2+n_4,\; b=2n_2, \; a=-2m_2-2n_2 \, ,
\label{m4n2}
\end{equation}
as recorded in Table \ref{tableXn}.

We notice that if expressions for solutions
\[ X_n^{1,2}=s_1 s_2 T_2^{-n_2} T_4^{n_4} (q=z,p=0)_{\mathsf{a}}
, \;\; Y_m =T_2^{-m_2} T_4^{m_4}(q=z,p=0)_{\mathsf{b}} \, , 
\]
were to be set equal, one could multiply both sides 
by $ s_3 $ and create divergence since $ s_3 T_4^{m_4}= T_3 ^{m_4} s_3 $ 
on  $(q=z,p=0)_{\mathsf{b}}$ is
forbidden. Thus we deal with degeneracy 
for parameters 
$\alpha_i^{(1,2)}$ from equation \eqref{alpha12}.

It appears that $X_{  n}^{1,2}$ is connected to
$X_{+, n, \mathsf{a}}^{1}$ via simple multiplication by 
the B\"acklund transformation
$T_1^{-1}T_2^{-1}s_2s_4$.
Consider namely
\begin{equation}
	\begin{split}
	(T_1^{-1}T_2^{-1}s_2s_4) \; \pi s_1&=	T_1^{-1}T_2^{-1} \pi s_1 s_3 s_1 
		=\pi T_4^{-1}T_1^{-1} s_3\\
		&=\pi s_3 T_3^{-1} T_1^{-1} = s_1 s_2 T_1^{-1} \, .
	\end{split}
\label{charada}
\end{equation}
Consequently, we obtain
\begin{equation}
	\begin{split}
		T_1^{-1}T_2^{-1}s_2s_4 X_{+, n, \mathsf{a}}^{1}&= s_1 s_2 T_1^{-1} T_2^{-n_2} T_4^{n_4}\,
		(q=z,p=0)_{\mathsf{a}}\\&=s_1 s_2  T_2^{-n_2} T_4^{n_4}\,
		(q=z,p=0)_{\mathsf{a}-1}= X_{  n,\mathsf{a}-1 }^{1,2}
		\, .
\label{charada1}
\end{split}
\end{equation}		
		Thus connecting $X_{+, n, \mathsf{a}}^{1}$ to
		$X_{  n,\mathsf{a}-1 }^{1,2}$.
The question is whether acting with $T_1^{-1}T_2^{-1}s_2s_4$ on 
$Y_{m} $ we obtain a valid solution ?
We calculate $T_1^{-1}T_2^{-1}s_2s_4 Y_{m} $ to find out if it is finite:
\begin{equation}
\begin{split}
&T_1^{-1}T_2^{-1}s_2s_4 T_2^{-m_2} T_4^{m_4}(q=z,p=0)_{\mathsf{b}}
= T_1^{-1}T_2^{-1} T_3^{-m_2} T_1^{m_4} s_2 s_4 (q=z,p=0)_{\mathsf{b}}\\
&= T_2^{-1} T_3^{-m_2+1} T_1^{m_4-1} 
\pi^{-1} s_4 s_1 s_4  (q=z,p=0)_{\mathsf{b}}
= T_3^{-m_2+1} T_1^{m_4-1} \pi^{-1} s_4 s_1 s_4 T_3^{-1} (q=z,p=0)_{\mathsf{b}}
\label{TonYa}
\end{split}
\end{equation}
because of a presence of $T_3^{-1}$ it follows that the above expression 
is infinite.  Thus the  B\"acklund transformation
$T_1^{-1}T_2^{-1}s_2s_4$ does not connect the pair of solutions
$X_{+, n, \mathsf{a}}^{1}, Y_m$ into
the pair of solutions $X_{  n,\mathsf{b} }^{1,2}, Y_m$ and the two
degeneracies connected with $X_{+, n, \mathsf{a}}^{1}$ and $X_{  n,\mathsf{b} }^{1,2}$
are independent. 
		
We will derive below few examples of
the degenerated pair of solutions for $X_{  n,\mathsf{b} }^{1,2}, Y_m$.

Let us set  $n_2=n_4 =n+1$ with $n \ge 0$.
We therefore find from the third row of Table  \ref{tableXn} :
\begin{equation}
	n_2= n_4=n+1 ,\; \mathsf{a}=-2(m_2+n+1), \; \mathsf{b}=2(n+1), m_4=2 (n+1),\;
\label{n2m2}
\end{equation}
for which we find $\alpha_i=2(n+1+m_2,-m_2,-2(n+1),n+2)$. 

\begin{exmp}
Setting   $m_2=0$ in relations \eqref{n2m2} we get the following
solutions:
\begin{equation}
s_1 s_2 T_2^{-n-1} T_4^{n+1} (q=z,p=0)_{\mathsf{a}=-2(n+1)} \, ,
\label{s1s2m2=0}
\end{equation}
and
\begin{equation}
T_4^{2(n+1)} (q=z,p=0)_{\mathsf{b}=2(n+1)} \, .  %
\label{T4m2=0}
\end{equation}
For the simplest case of $n=0$ and $m_2=0$ we obtain :
\[n_2= n_4=1 ,\; a=-2, \; b=2, m_4=2,
\]
with solutions 
\[ 
s_1 s_2 T_2^{-1} T_4^1 (q=z,p=0)_{\mathsf{a}=-2}=
\left( q= -\frac{2(4+z^2)}{z(2+z^2)}, \; p=
\frac{z^5}{(4 z^2+z^4+8)} \right) \, ,
\]
and
\[
 T_4^2 (q=z,p=0)_{\mathsf{b}=2}= \left( q= \frac{z^2-4}{z}, p=0 
\right) \, ,
\]
Both expressions are solutions of PV Hamilton equations with 
$\alpha_i=(2,0,-4,4)$.
For $n=1$, and still with $m_2=0$, we get from expression
\eqref{s1s2m2=0} :
\[ \begin{split}
q &= -4\frac{ (48 + 24 z^2 + 6 z^4 + z^6) (1152 + 768 z^2 + 216 z^4 +
24 z^6 + z^8)}{z (192 + 72 z^2 + 12 z^4 + z^6)(192 + 192 z^2 + 96 z^4
+ 16 z^6 + z^8)}\, ,\\
p&=\frac{z^7 (192 + 72 z^2 + 12 z^4 + z^6)^2}{(
48 + 24 z^2 + 6 z^4 + z^6) (9216 + 9216 z^2 + 4608 z^4 + 1536 z^6 +
264 z^8 + 24 z^{10} + z^{12})}\, ,
\end{split}
\]
and from expression \eqref{T4m2=0} :
\[ q= \frac{(-2 + z) (2 + z) (-12 + z^2)}{
z (-8 + z^2)}, \;\; p=0 \, ,
\]
both satisfying the PV equation with $\alpha_i=(4, 0,  -8, 6)$.   

\end{exmp}
\begin{exmp}
Setting $m_2=1$ in relations \eqref{n2m2}
we find  $\alpha_i=2(n+2,-1,-2(n+1),n+2)$ with
$\mathsf{a}=-2(n+2), \; \mathsf{b}=2(n+1), m_4=2 (n+1)$
for which we find for $n=0$:
\[
q= -
4 \frac{(6 + z^2)}{
z(4 + z^2)}, \; p= \frac{z (8 + 4 z^2 + z^4)
}{
24 + 8 z^2 + z^4}\, ,
\]
using $X_n^{1,2}=s_1 s_2 T_2^{-1} T_4^{1} (q=z,p=0)_{\mathsf{a}=-4}$
and
\[q= \frac{
24 - 8 z^2 + z^4
}{z
(-2 + z) (2 + z)} , \; p=\frac{2 z}{6 - z^2} \, ,
\]
using  $T_2^{-1} T_4^{2}(q=z,p=0)_{\mathsf{b}=2}$. Both solutions to
the PV equation share the same
$\alpha_i= (4,-2, -4,4)$.

\end{exmp}

\section{Discussion and Outlook}
\label{discussion}

First let us remark for completeness that although  discussion in this paper 
was given mainly for solutions
derived from the seed solution $(q=z,p=0)_{\mathsf{a}}$ it can
straightforwardly be extended to other $\pi$-variants defined in
\eqref{wata2}-\eqref{wata6} by applying transformations $M \to \pi M
\pi^{-1}$ and $M \to \pi^{-1} M
\pi$ for $M=s_1 s_2, s_3 s_4$ used above.

Let us comment on how our results compare to general expectations 
regarding degeneracy.

For linear systems, as the ones encountered in for example Quantum
Mechanical systems, as $3D$ Hydrogen atom, the presence of degeneracy
signals presence of symmetry (e.g. $SO(4) $ symmetry for the Hydrogen atom)
and symmetry is acknowledged to cause  degeneracy.
Conversely, for Quantum Mechanics with eigenvectors associated with one eigenvalue
one can define a symmetry operation that rotates eigenvectors into each other
thus connecting symmetry to degeneracy. 
Note however that the system we consider here is nonlinear so any linear
combination of degenerated solutions is not expected to be a solution.
Rather than looking for a symmetry explanation for existence of
degeneracy we point out that exclusion rules for a class of solution
derived from the seed solution \eqref{solution1}, as well as the fact that
translation operators in such system do not span a complete set of
solutions as technical but fully
complete explanations for why we encountered degeneracy for the PV equation.

Our discussion clearly indicates that degeneracy will exist for 
all dressing chains of even periodicity because of existence of 
exclusion rules for classes of seed solutions,
see \cite{AGLZ} for discussion $N=6$ periodic dressing chain. 
It will be interesting to investigate how many solutions will share
their $\alpha$'s in case of higher periodic dressing chains. 
It would be important to find if all the solutions with degeneracy are
described by special hypergeometric functions that as we have seen 
in  \cite{AGLZ} takes place for solutions generated by powers of 
$T_2$ and $T_4$ from the seed solution $(q=z,p=0)_{\mathsf{a}}$, especially in 
view of a role played by B\"acklund transformations $s_i$ that
supplement the  translation
operators when deriving solutions with degeneracy.

Next we would like to comment on relation between $s_i$ and translation
operators as this relation turned out to be crucial for presence of
degeneracy.
By definition translation operators are expressed 
by B\"acklund transformations $s_i, i=1,2,3$ and the automorphism $\pi$.
We encountered in subsection \ref{subsection:item1&2}  an inverse
relation with $s_i$ B\"acklund transformations' action on seed solutions
\eqref{solution2} realized as action of translation operators when acting 
on the same seed solutions with $\mathsf{a}$ of an appropriate parity.

We will now show that  if one introduces square-roots
of  products of translation operators defined as well-defined 
B\"acklund transformations then the inverse relations can be
established that reproduce $s_i$ in terms of translation operators and
their square roots.

We first  consider a quartic product of two B\"acklund transformations $s_is_j$. 
One can express such products of $s_is_j$ in the following way  
\begin{equation}
 (s_{i-1} \pi s_i )^2 = T_i T_{i+2}, \quad (s_{i-1} \pi s_{i+1} )^2 
=  T_{i} T_{i+1} \, , 
\label{squares}
\end{equation}
which can then be used to define
\begin{equation}
 (T_i T_{i+2})^{1/2}= s_{i-1} \pi s_i  , \quad 
 (T_{i}T_{i+1})^{1/2}= s_{i-1} \pi s_{i+1}  \, ,
\label{squares1}
\end{equation}
as B\"acklund transformations.

For $A_3^{(1)}$ extended affine Weyl group
there are two  square-roots of the type $(T_i T_{i+2})^{1/2}$,for
$i=1$ and $i=2$.
They transform $(\alpha_1, \alpha_2, \alpha_3, \alpha_4)$
as follows
\begin{equation}
\begin{split}
(T_1 T_{3})^{1/2} (\alpha_1, \alpha_2, \alpha_3, \alpha_4)&=
(2-\alpha_2, -\alpha_1, 2- \alpha_4,-\alpha_3)\, ,\\
(T_2 T_{4})^{1/2} (\alpha_1, \alpha_2, \alpha_3, \alpha_4)&=
(-\alpha_4, 2-\alpha_3, - \alpha_2,2-\alpha_1)\, , 
\end{split}
\label{TiTi+2sqr}
\end{equation}
which when applied twice successively yield the results for 
$T_1 T_{3}$ and $T_2 T_{4}$ that fully agree with relations \eqref{Tialpha}.

It follows from equations  \eqref{TiTi+2sqr}
that for  $\alpha_i=(\mathsf{a}, 1-\mathsf{a}, \mathsf{a}, 1-\mathsf{a})$
the transformation $(T_i T_{i+2})^{1/2}$ shifts the parameter 
$\mathsf{a}$ : $\mathsf{a} \to \mathsf{a}-(-1)^i$, furthermore it can be
shown that they leave $q=p=z/2$ 
of definition \eqref{solution2} unchanged, which is summarized as
follows:
\begin{align} 
(T_1T_3)^{1/2} (z/2,z/2)_{\mathsf{a}}&=
(z/2,z/2)_{\mathsf{a}+1}\, ,\label{T1T3halfa}\\
(T_2 T_4)^{1/2} (z/2,z/2)_{\mathsf{a}}&=
(z/2,z/2)_{\mathsf{a}-1} \, .
\label{T2T4halfa}
\end{align}

For  square-roots of the type $(T_i T_{i+1})^{1/2}$
we consider an example of $(T_2 T_{3})^{1/2}$
to find that it transforms $(\alpha_1, \alpha_2, \alpha_3, \alpha_4)$
as follows
\begin{equation}
(T_2 T_{3})^{1/2} (\alpha_1, \alpha_2, \alpha_3, \alpha_4)=
(-\alpha_3-\alpha_4,2 -\alpha_2, \alpha_2+ \alpha_3,\alpha_4)\, .
\label{T2T3sqr}
\end{equation}
Applying this transformation twice yields
$(T_2 T_{3})^{2/2} (\alpha_1, \alpha_2, \alpha_3, \alpha_4)=
(\alpha_1-2, \alpha_2, 2+\alpha_3, \alpha_4)$ in agreement
with  relations \eqref{Tialpha}.

We will now show how to express a single
B\"acklund transformation $s_i$ in terms of a product and square roots of
translation operators.
To achieve this we combine the identities \eqref{squares1} with expressions \eqref {nextsi} 
and using that $\pi s_{i+2} s_{i+1}=s_{i+3} \pi s_{i+1}=
s_{i-1} \pi s_{i+1}$
we find the following expressions  for $s_i$:
\begin{equation}
s_i =  T_i^{-1} (T_i T_{i+1})^{1/2} , \; i=1,2,3,4
\label{sisquare}
\end{equation}
One can show that $(T_i T_{i+1})^{1/2} T_i=T_{i+1} (T_i T_{i+1})^{1/2}$
from which $s_i^2=1$ easily follows.

As an  example  we find $T_1T_4 = (s_3 \pi s_1)^2$ that can be used to
define $(T_1 T_4)^{1/2} = s_3 \pi s_1$. It easy to show an identity
$\pi s_2 s_1 = s_3 \pi s_1 =T_4 s_4$ from which we obtain
\[ s_4 = T_4^{-1} ( T_4 T_1)^{1/2} \, . \]

We will return elsewhere to derivation of $A^{(1)}_3$ group relations
from natural commutation relations for square-roots of translations
and translation operators.

\vspace{5mm}

{\bf Acknowledgments}
This study was financed in part  by the Coordena\c{c}\~{a}o de Aperfei\c{c}amento de Pessoal de N\'ivel Superior - Brasil (CAPES) - Finance Code 001
(G.V.L.) and by CNPq and FAPESP (J.F.G. and A.H.Z.).

\end{document}